\documentclass[fleqn,usenatbib]{mnras}
\usepackage[T1]{fontenc}
\usepackage{ae,aecompl}

\usepackage{graphicx}
\usepackage{amsmath}	
\usepackage{amssymb}

\usepackage{times}
\usepackage{hyperref}

\title[3C~279 in 2008--2018]{Multiwavelength behaviour of the blazar 3C~279: decade-long study from $\gamma$-ray to radio}

   \author[V.M.\,Larionov et al.]
	{\parbox{\textwidth}{V.M.\,Larionov,$^{1,2}$\thanks{e-mail {\tt v.larionov@spbu.ru}}
S.G.\,Jorstad,$^{1,3}$
A.P.\,Marscher,$^{3}$
M.\,Villata,$^{4}$
C.M.\,Raiteri,$^{4}$
P.S.\,Smith,$^{5}$
I.\,Agudo,$^{6}$
S.S.\,Savchenko,$^{1}$
D.A.\,Morozova,$^{1}$
J.A.\,Acosta-Pulido,$^{7,8}$
M.F.\,Aller,$^{9}$
H.D.\,Aller,$^{9}$
T.S.\,Andreeva,$^{10}$
A.A.\,Arkharov,$^{2}$
R.\,Bachev,$^{11}$
G.\,Bonnoli,$^{12}$
G.A.\,Borman,$^{13}$
V.\,Bozhilov,$^{14}$
P.\,Calcidese,$^{15}$
M.I.\,Carnerero,$^{4}$
D.\,Carosati,$^{16,17}$
C.\,Casadio,$^{18}$
W.-P.\,Chen,$^{19}$
G.\,Damljanovic,$^{20}$
A.V.\,Dementyev,$^{1}$
A.\,Di Paola,$^{21}$
A.\,Frasca,$^{22}$
A.\,Fuentes,$^{6}$
J.L.\,G\'omez,$^{6}$
P.\,G\'onzalez-Morales,$^{7,8}$
A.\,Giunta,$^{23}$
T.S.\,Grishina,$^{1}$
M.A.\,Gurwell,$^{24}$
V.A.\,Hagen-Thorn,$^{1}$
T.\,Hovatta,$^{25,26}$
S.\,Ibryamov,$^{27}$
M.\,Joshi,$^{3}$
S.\,Kiehlmann,$^{28}$
J.-Y.\,Kim,$^{18}$
G.\,N.\,Kimeridze,$^{29}$
E.N.\,Kopatskaya,$^{1}$
Yu.A.\,Kovalev,$^{30}$
Y.Y.\,Kovalev,$^{30,31,18}$
O.M.\,Kurtanidze,$^{29,32,33}$
S.O.\,Kurtanidze,$^{29}$
A.\,L\"ahteenm\"aki,$^{26,34}$
C.\,L\'azaro,$^{7,8}$
L.V.\,Larionova,$^{1}$
E.G.\,Larionova,$^{1}$
G.\,Leto,$^{22}$
A.\,Marchini,$^{12}$
K.\,Matsumoto,$^{35}$
B.\,Mihov,$^{11}$
M.\,Minev,$^{14}$
M.G.\,Mingaliev,$^{36,37}$
D.\,Mirzaqulov,$^{38}$
R.V.\,Mu\~noz Dimitrova,$^{11}$
I.\,Myserlis,$^{18}$
A.A.\,Nikiforova,$^{1,2}$
M.G.\,Nikolashvili,$^{29}$
N.A.\,Nizhelsky,$^{36}$
E.\,Ovcharov,$^{14}$
L.D.\,Pressburger,$^{3}$
I.A.\,Rakhimov,$^{10}$
S.\,Righini,$^{39}$
N.\,Rizzi,$^{40}$
K.\,Sadakane,$^{35}$
A.C.\,Sadun,$^{41}$
M.R.\,Samal,$^{19,42}$
R.Z.\,Sanchez,$^{22}$
E.\,Semkov,$^{43}$
S.G.\,Sergeev,$^{13}$
L.\,A.\,Sigua,$^{29}$
L.\,Slavcheva-Mihova,$^{11}$
P.\,Sola,$^{7}$
Yu.V.\,Sotnikova,$^{36}$
A.\,Strigachev,$^{11}$
C.\,Thum,$^{44}$
E.\,Traianou,$^{18}$
Yu.V.\,Troitskaya,$^{1}$
I.S.\,Troitsky,$^{1}$
P.G.\,Tsybulev,$^{36}$
A.A.\,Vasilyev,$^{1}$
O.\,Vince,$^{20}$
Z.R.\,Weaver,$^{3}$
K.E.\,Williamson,$^{3}$
G.V.\,Zhekanis$^{36}$
}
}
\date{Accepted 2020 January 8. Received 2019 December 12; in original form 2019 October 3.}
\pubyear{2020}	
\begin{document}
\label{firstpage}
\pagerange{\pageref{firstpage}--\pageref{lastpage}} 
\maketitle

\begin{abstract}%
 We report the results of decade-long (2008--2018) $\gamma$-ray to 1~GHz radio monitoring of the blazar 3C~279, including GASP/WEBT, \emph{Fermi} and \emph{Swift} data, as well as  polarimetric and spectroscopic data. The X-ray and $\gamma$-ray light curves correlate well, with no delay $\ga 3$ hours, implying general co-spatiality of the emission regions. The $\gamma$-ray-optical flux-flux relation changes with activity state, ranging from a linear to a more complex dependence. The behaviour of the Stokes parameters at optical and radio wavelengths, including  43~GHz VLBA images, supports  either a predominantly helical magnetic field or motion of the radiating plasma along a spiral path. Apparent speeds of emission knots range from 10 to 37c, with the highest values requiring bulk Lorentz factors close to those needed to explain $\gamma$-ray variability on very short time scales. The \ion{Mg}{II} emission line flux in the `blue' and `red' wings correlates with the optical synchrotron continuum flux density, possibly providing a variable source of seed photons for inverse Compton scattering. In the radio bands we find progressive delays of the most prominent light curve maxima with decreasing frequency, as expected from the frequency dependence of the $\tau=1$ surface of synchrotron self-absorption. The global maximum in the 86~GHz light curve becomes less prominent at lower frequencies, while a local maximum, appearing in 2014, strengthens toward decreasing frequencies, becoming pronounced at $\sim5$~GHz. These tendencies suggest different Doppler boosting of stratified radio-emitting zones in the jet.
\end{abstract}

%
%
\section{INTRODUCTION}
\label{intro}
The spectral energy distribution of the blazar subclass of active galactic nuclei -- which includes flat-spectrum radio quasars, FSRQ, and BL~Lac-type objects -- is dominated 
by highly variable nonthermal emission from relativistic jets that are viewed within
several degrees of the jet axis.
The FSRQ 3C~279 \citep[redshift $z=0.538$,][]{Burbidge1965}; now commonly accepted $z=0.5362$, \citet{Marziani1996}, considered a prototypical 
blazar, exhibits pronounced variations of flux from radio to
$\gamma$-ray frequencies (by $>5$ mag in optical bands) and strong, variable optical polarization as high as 45.5 per cent, observed at $U$ band \citep{Mead1990}. It has been
intensely observed in many multiwavelength campaigns designed to probe the physics of the 
high-energy plasma responsible for the radiation. In one such campaign in 1996,
3C~279 was observed to vary rapidly during a high flux state by the EGRET detector of the 
{\it Compton} Gamma Ray Observatory and at longer wavelengths \citep{Wehrle1998}.
The $\gamma$-ray maximum coincided with X-ray flare with no time lag within 1 day. Since 
the launch of \emph{Fermi}, numerous observational campaigns have been undertaken in order
to trace the variability of 3C~279 on different time scales and over an as wide a wavelength
range as possible \citep[e.g.,][]{Bottcher2007, Larionov2008, Hayashida2012, Pittori2018}.
Although such multi-epoch campaigns, each extending over relatively a brief time interval, 
have led to improvements in our knowledge about the blazar, they have
proven insufficient to discover patterns in the complex behaviour of 3C~279 that relate to 
consistent physical aspects of the jet.

Polarimetric studies of 3C~279 are still not as numerous as photometric ones. During a 2007 dedicated observational campaign \citep{Larionov2008}, the optical and 7-mm EVPAs (electric vector position angles) rotated simultaneously, suggesting co-spatiality of the 
optical and radio emission sites. The same study found that the slope of the variable source of the optical (synchrotron) spectrum did not change, despite large ($\sim 3$ mag) variations in brightness.  In a recent work by \citet{Rani2018}, an anticorrelation was found between the optical flux density and the degree of optical polarization. \citet{Kiehlmann2016} carried out a detailed analysis of 3C~279 polarimetric behaviour in order to distinguish between `real' and `random walk' rotations of the EVPA, finding that a smooth optical EVPA rotation by $\sim 360\degr$ during a flare was inconsistent with a purely stochastic process.

\citet{Punsly2013} has reported a prominent asymmetry (strong `red wing') of
the \ion{Mg}{ii} line based on examination of three spectra of 3C~279 obtained at Steward Observatory. The origin of the red wing is unknown, with explanations ranging
from gravitational and transverse redshifts within 200 gravitational radii from the central black hole~\citep[see, e.g.,][]{Corbin1997}, reflection off optically thick, out-flowing clouds on the far side of the accretion disc, or transmission through inflowing gas on the near side of the accretion disc.

High-resolution observations with the Very Long Baseline Array (VLBA) have demonstrated that $\gamma$-bright 
blazars possess the most highly relativistic jets among compact flat-spectrum radio sources 
\citep[e.g.,][]{Jorstad2001b}. This is inferred from comparison of kinematics of disturbances (knots) in parsec-scale jets of AGN that are strong $\gamma$-ray 
emitters versus those that are weak or undetected. 
Reported apparent speeds in 3C~279 have ranged from
4$c$ to 22$c$ \citep[e.g.,][]{Jorstad2004, Rani2018, Lister2019}.
Based on a detailed study of knot trajectories, 
\citet{Qian2019} have suggested that the jet precesses, as could be caused by a binary black hole system in the nucleus.
From an analysis of both the apparent speeds and the time scales of flux decline of individual knots in 3C~279, \citet{Rani2018} derived a Lorentz factor of the jet
flow $\Gamma \ge 22.4$, and a viewing angle  $\Theta \le 2\fdg6$.

An analysis of times of high $\gamma$-ray flux and superluminal ejections of $\gamma$-ray blazars indicates a statistical connection between the two types of events \citep{Jorstad2001a,Jorstad2016,Rani2018}.

Given the complexity of the time variability of non-thermal emission in blazars, a more complete observational
dataset than customarily obtained might extend the range of conclusions that can be drawn concerning
jet physics. \citet{Chatterjee2008} analysed a decade-long (1996--2007) dataset containing
radio (14.5 GHz), single-colour ($R$-band) optical, and X-ray light curves, as well as VLBA images at 43 GHz, of 3C~279. They
found strong correlations between the X-ray and optical fluxes, with time delays that change from X-ray leading
optical variations to vice-versa, with simultaneous variations in between. Although the radio variations
lag behind those in these wavebands by more than 100 days on average, changes in the jet direction on time scales
of years are correlated with -- and actually lead -- long-term variations at X-ray 
energies. 

The current paper extends these observations
to include multi-colour optical, ultraviolet (UV), and near-infrared (NIR) light curves, as well as linear polarization in the optical
range, radio (350~GHz to 1~GHz) single-dish data, and 43~GHz VLBA images. We analyse optical spectroscopic observations performed at the Discovery Channel Telescope (DCT) telescope of Lowell Observatory. We also use open-access data from the \emph{Swift} (UVOT and XRT) and \emph{Fermi }satellites.  The radio-to-optical data presented in this paper have been acquired during a multifrequency campaign organized by the Whole Earth Blazar Telescope (WEBT).\footnote{{\tt http://www.oato.inaf.it/blazars/webt/} \citep[see e.g.][]{Bottcher2005, Villata2006, Raiteri2007}.}  For completeness, our study includes data from the 2006--2007 WEBT campaigns presented by \citet{Bottcher2007, Larionov2008,Hayashida2012}, and \citet{Pittori2018}.

Our paper is organized as follows:
Section~\ref{sect:observ} outlines the procedures that we used to process and analyse the data. Section~\ref{color_evolution} displays and analyses
the multifrequency light curves, while optical spectra are presented and discussed in \S\ref{spectra}. In \S\ref{sect:sed} we present the spectral energy distribution from radio to $\gamma$-ray frequencies constructed at six different flux states. 
In \S\ref{correlations} we derive and discuss the time lags between variations in different colour bands. Section \ref{vlba} describes the kinematics of the radio jet as
revealed by the VLBA images, and \S\ref{interpretation} relates changes in the structure
of the radio jet to the multiwavelength flux variability. Section~\ref{pol_behaviour} gives the results of the optical and radio polarimetry. In \S\ref{conclusions} we
discuss the implications of our observational results with respect to the physics of the jet in 3C~279. Abbreviation $TJD$ holds for $JD-2400000.0$. We use parameters for distances and cosmology $H_0= 73.0, \Omega_{\mathrm matter} = 0.27, \Omega_{\mathrm vacuum} = 0.73$.

\section{OBSERVATIONS AND DATA REDUCTION}
\label{sect:observ}

\subsection{Optical and near-infrared photometry}

\begin{table}
\begin{minipage}[t]{0.9\columnwidth}

\caption{Ground-based observatories participating in this work.}
\label{obs}
\begin{centering}
\renewcommand{\footnoterule}{}  
\begin{tabular}{l r c}
\hline\hline
Observatory    & Country     & Bands\\
\hline
\multicolumn{3}{c}{\it Optical}\\
Abastumani    &  Georgia         & $R$                \\
Belogradchik  & Bulgaria         & $B, V, R, I$                \\
Calar Alto$^a$  & Spain        & $R$          \\
Campo Imperatore (Schmidt) &  Italy        & $B, V, R$          \\
Catania         &  Italy         & $R$          \\
Crimean AZT-8 (AP7 \& ST-7$^b$) &  Russia       & $B, V, R, I$       \\
Lowell (Perkins$^b$ \& DCT$^c$)    &  USA&   $B, V, R, I$  \\
Lulin           & Taiwan          & $R$                \\
Mt.Maidanak   & Uzbekistan        & $B, V, R, I$       \\
New Mexico Skies (iTelesope) & USA &   $V, R$  \\
Osaka Kyoiku     &  Japan      & $B, V, R, I$                \\
Roque (KVA \& LT)      &   Spain       & $R$                \\
Rozhen (200 cm \& 50/70 cm)       & Bulgaria      & $U, B, V, R, I$    \\
San Pedro Martir          &  Mexico        & $R$                \\
Siena        &  Italy         & $R$          \\
Sirio         &  Italy         & $R$          \\
St.\ Petersburg$^b$   &  Russia        & $B, V, R, I$       \\
Steward (Kuiper, Bok \& MMT)$^d$  & USA & $R$ \\
Teide (IAC80 \& STELLA-I)   & Spain      & $R$       \\
Tijarafe   & Spain      & $R$       \\
Valle d'Aosta      & Italy        & $R, I$             \\
Vidojevica (140 \& 60 cm) & Serbia & $B, V, R, I$ \\
\hline
\multicolumn{3}{c}{\it Near-infrared}\\
Campo Imperatore (AZT-8)  &  Italy       & $J, H, K$          \\
Teide (TCS)   & Spain      &  $J, H, K$        \\

\hline
\multicolumn{3}{c}{\it Radio}\\
Mauna Kea (SMA) & USA     & 230, 345 GHz      \\
Medicina        & Italy & 5, 8 GHz \\
Mets\"ahovi  & Finland     & 37 GHz              \\
Noto    &  Italy & 5, 8, 43 GHz  \\
OVRO             &  USA         & 15 GHz              \\
Pico Veleta (IRAM) & Spain &  86, 229 GHz \\
RATAN-600 & Russia    & 1, 2,  5,  8, 11, 22 GHz      \\
Svetloe & Russia  & 5, 8 GHz \\
VLBA  &  USA   &  43 GHz \\
UMRAO  &  USA               & 4.8, 8, 14.5 GHz      \\
\hline
\end{tabular}
\end{centering}
$^a$ -- MAPCAT project: http://www.iaa.es/∼iagudo/research/MAPCAT\\
$^b$ -- photometry and polarimetry \\
$^c$ -- spectroscopy  \\
$^d$ -- spectropolarimetry
\end{minipage}

\end{table}

Observations of 3C~279 under the GASP-WEBT program
~\citep[see e.g.,][]{2008A&A...481L..79V, 2009A&A...504L...9V} were performed in 
2008--2018 at optical, near-infrared (NIR), and radio bands at 32 observatories,
listed in Table~\ref{obs}.

We performed photometry in optical bands, as described in \cite{Raiteri1998}, and in NIR bands, as detailed in \cite{Gonzalez-Perez2001}. 
Following standard WEBT procedures, we compiled and `cleaned' the optical light curves 
\citep[see, e.g.,][]{Villata2002}. As needed, we applied systematic corrections (mostly from effective wavelengths deviating from those of standard $BVR_CI_C$
bandpasses) to the data from some of the telescopes, in order to adjust the calibration
so that the data are consistent with those obtained via standard instrumentation and 
procedures. The resulting offsets are generally less than 0.01--0.03 mag in $R$ band.

We have corrected the optical and NIR data for Galactic extinction using values for each 
filter reported in the NASA Extragalactic Database
(NED)\footnote{\url{http://ned.ipac.caltech.edu/}} \citep{Schlafly2011}.
The magnitude to flux conversion adopted the coefficients of \cite{Mead1990}.

\subsection{\textit{Swift} observations}
\subsubsection {Optical and ultraviolet data}
The quasar 3C~279 was observed with the $UVOT$ instrument of the Neil Gehrels \emph{Swift} Observatory in all 
6 available filters, $UVW2$, $UVM2$, $UVW1$, $U$, $B$, and $V$.
The $UVOT$ data of 3C~279 were downloaded from the \emph{Swift} archive and reduced using 
the HEAsoft version 6.24 \emph{Swift} sub-package and corresponding HEASARC 
calibration database (CALDB). If necessary, aspect correction was performed using the \textsc{uvotunicorr} task and  
all exposures within an observation were summed using the \textsc{uvotimsum} task. A magnitude was derived using 
\textsc{uvotsource}
with a 5 arcsec radius aperture centred on the object and a 20 arcsec radius aperture located in a source-free area away from 
the object for the background. The result was tested for a large coincidence-loss correction factor. Observations with a 
large magnitude error or an exposure time of less than 40 seconds were discarded.
Galactic extinction correction in the ultraviolet (UV) bands was performed using the interstellar extinction curve of \cite{Fitzpatrick1999}
with $R_V = 3.1$. The corrections of the $U$, $B$, and $V$ band data were made according to the \cite{Schlafly2011} values, as listed in 
the NASA Extragalactic Database ($A_U=0.124, A_B=0.104, A_V=0.078$).
All of the derived parameters are given in Table~\ref{swift_calib}.

\begin{table}
\caption{\bf Swift calibrations used for  3C~279 analysis.}
\label{swift_calib}
\begin{tabular}{c | c | c | c | c | c | c |}
\hline

 Bandpass & v & b &  u & uw1 & um2 & uw2 \\
\hline

$\lambda$, \AA & 5427 & 4353 & 3470 & 2595 & 2250 & 2066 \\
\hline
$A_\lambda$, mag & 0.078 & 0.104& 0.124 & 0.175 & 0.265 & 0.251 \\
\hline
conv. factors  &2.603 & 1.468 & 1.649 & 4.420 & 8.372 & 5.997 \\
\hline
\end{tabular}\\
\flushleft{Note -- Units of count rate to flux conversion factors are $10^{-16}{\rm erg}\,{\rm cm^{-2}s^{-1}}$\AA$^{-1}$.}

\end{table}

\subsubsection{X-ray data}
We downloaded data obtained for 3C~279 with the X-Ray Telescope (XRT) of the Neil Gehrels \emph{Swift} Observatory 
from the \emph{Swift} archive, which covers the period from 2006 January to 2018 June.  All observations were performed
in Photon Counting (PC) mode, with an exposure time between 1-3~ksec, except for two cases 
when the quasar was in an active state and several exposures as short as 200 seconds were carried out during
a given day.  The XRT data were reduced using the HEAsoft version 6.24 \emph{Swift} sub-package and corresponding HEASARC CALDB. The FTOOLS task \textsc{xrtpipeline} was run to clean the data and to create an exposure 
file used in the task \textsc{xrtmkarf}, which generates an Ancillary Response Function file for input into the spectral 
fitting program \textsc{Xspec}. The source and background images and spectra were extracted using the task \textsc{Xselect}. 
For the source, photons were counted over a circular region of 20-pixel radius centred on the object's coordinates,
while for the background region a larger annulus was used, with inner and outer radii of
75 and 100 arcsec, correspondingly, centred on the source and selected to avoid 
any contaminating sources. In preparation for running \textsc{Xspec},     
we grouped the energy channels via the \textsc{grppha} tool such that each energy bin contained at least one photon, as recommended in the \textsc{Xspec} manual \citep{Arnaud1996} for Cash statistics, which we used to evaluate the goodness of fit \citep{Cash1979}. Spectra of 3C~279 from 0.3 to 10 keV were fit within Xspec using an absorbed power law model, with the $n_{\mathrm H}$ parameter fixed at $2.21\cdot 10^{20}$ cm$^{-2}$ \citep{Dickey1990}. 
Each epoch that suffered from pile-up ($>  0.5$ cnts/sec for PC data ) was re-examined as outlined in 
\url{ http://www.swift.ac.uk/analysis/xrt/pileup.php}. This resulted in creation of a new annular source region, with the 
size of the inner ring determined by the King function \citep{Moretti2005}. 

\subsection{\emph{Fermi} LAT observations}
\label{sec:LAT}
The $\gamma$-ray data were obtained with the {\emph Fermi} (LAT),
which observes the entire sky every 3 hours at energies of 20~MeV--300~GeV \citep{Atwood2009}. The construction of the $\gamma$-ray light curve and the time-dependent spectral energy distribution (SED)
was based on \emph{Fermi} Large Area Telescope observations at 0.1--200~GeV, obtained via the open-access
mission website\footnote{\url{https://fermi.gsfc.nasa.gov/}}. To obtain the light curve, we used the standard
unbinned  likelihood analysis pipeline of the \emph{Fermi} Science Tools {\sc v10r0p5} package with the instrument
response function {\sc p8r2\_v6}, Galactic diffuse emission model {\sc gll\_iem\_v06}, and isotropic background
model  {\sc iso\_pr2\_source\_v6\_v06}. We used the adaptive binning technique described in \cite{Larionov2017}
to break the entire data time range into integration bins of variable length, such that high-flux periods were
split into shorter bins than were other time spans. This approach allows us to have very fine (down to $0\fd25$) binning
for active states while maintaining a significant signal level (maximum-likelihood test statistic $TS \ge 10$) during low-flux periods.

To study $\gamma$-ray SED variations over time, we performed a binned likelihood analysis, which allows
one to find a flux value over a set of energy bins, in contrast to an unbinned analysis, which can infer only
the total photon flux over all energies. We split the entire time range into 50 time bins of roughly the same total
flux. To do so, we integrated the light curve over time using fixed values of spectral parameters to estimate the total
energy received from the object in $\gamma$-rays. We then divided the light curve into 50 sub-sections, each
containing 2.0 per cent of the total energy. The number 50 was adopted empirically, such that the flux integrated over the resulting bins
had a high enough significance level to run the binned likelihood analysis: the number of logarithmically distributed
energy bins with detected signal was at least 12 in every time bin. The shortest time bin we obtained was 2.75 days,
while the longest was almost 312 days. For running the binned analysis, we used the {\sc fermipy} Python wrapper~\citep{Wood2017}
around the standard \emph{Fermi}\,ST package.

\subsection{Radio observations}
The University of Michigan Radio Astronomy Observatory (UMRAO) total flux density observations presented here 
were obtained at frequencies of 4.8, 8.0, and 14.5 GHz
($\lambda =$ 6.2, 3.8, and 2.0~cm) with the UMRAO 26-m paraboloid antenna as part of a long-term extragalactic 
variable source monitoring program \citep{1985ApJS...59..513A}.  The adopted flux density scale is based on 
\citet{1977A&A....61...99B} and uses Cassiopeia A (3C 461) as the primary standard. In addition to observations 
of Cas A, observations of nearby secondary flux density calibrators were interleaved with observations of the 
target sources, typically every 1.5 to 2 hours, in order to verify the stability of the antenna gain and the 
accuracy of the telescope pointing. Each daily-averaged observation of 3C~279 consisted of a series of 8--16 
individual measurements obtained over 25--45 minutes, and observations were carried out when the source was 
within 3 hours of the meridian. 

Observations with the RATAN-600 radio telescope, operated by the Special Astrophysical 
Observatory (SAO) of the Russian Academy of Science (RAS), were carried out
in transit mode \citep{1972IzPul.188....3K,1979S&T....57..324K,1993IAPM...35....7P}.
The flux density measurements were obtained at 6 frequencies -- 22, 11, 7.7, 4.8, 2.3,
and 1.2 (or 0.97)~GHz ($\lambda =$ 1.38, 2.7, 3.9, 6.2,
13 and 24 (or 31)~cm) -- over several minutes per object. All continuum
receivers are total-power radiometers with square-law detection. The data are
registered using a regular universal registration system based on the
hardware-software subsystem ER-DAS (Embedded Radiometric Data Acquisition System)
\citep{2011AstBu..66..109T}. The data reduction procedures and main parameters of the antenna and radiometers are
described in \cite{1999A&AS..139..545K,2014A&A...572A..59M} and 
\cite{2016AstBu..71..496U}.

Observations at 5, 8 and 43 GHz were performed with the Medicina and Noto radio telescopes; a description of data reduction and analysis can be found in \citet{DAmmando2012}. 

3C~279 was also observed at 15\,GHz as part of high-cadence blazar monitoring program using the Owens Valley Radio Observatory (OVRO) 40-m Telescope \citet{Richards2011}. Observations were performed approximately twice per week. 3C~286 was used as the primary flux density calibrator with the scale set to 3.44 Jy following \citet{Baars1977}.

In the period from March 8, 2018 to January 2, 2019, radio observations at 4.8 and 8.5~GHz ($\lambda =$ 6.2 and 3.5~cm) were obtained with the 32~m diameter, fully-steerable radio telescope RT-32 at the Svetloe Observatory of the Institute of Applied Astronomy of the RAS.  The source 3C~295 was used as the primary calibration standard, and the secondary standard was DR21. The average measurement errors are estimated as 2 per cent (4.8~GHz, 61 points) and 1 per cent (8.5~GHz, 23 points). 

The short millimetre wavelength data presented in this paper were obtained under the POLAMI (Polarimetric Monitoring of AGN at Millimetre Wavelengths) Program\footnote{http://polami.iaa.es}  \citep{Agudo2018a,Agudo2018b,Thum2018}. POLAMI is a long-term program to monitor the polarimetric properties (Stokes I, Q, U, and V) of a sample of about 40 bright AGN at 3.5 and 1.3 millimetre wavelengths with the IRAM 30m Telescope near Granada, Spain. The program has been running since October 2006, and it currently has a time sampling of $\sim$2 weeks. The XPOL polarimetric observing setup,
described in \citet{Thum2008}, has been routinely used since the start of the program. The reduction and calibration of the POLAMI data presented here are described in detail in \citet{Agudo2010,Agudo2014,Agudo2018a}. 

Observations at 230 and 350 GHz were obtained as part of the long-term, ongoing monitoring program of mm-wave calibrator sources using the Submillimetre Array, near the summit of Mauna Kea, Hawaii  (see http://sma1.sma.hawaii.edu/callist/callist.html).

\subsection{Very long baseline interferometry}
Since 2007 June, the quasar 3C~279 has been monitored approximately monthly by the Boston University (BU) 
group with the Very Long Baseline Array (VLBA) at 43 GHz within a sample of gamma-ray and radio bright blazars 
(the VLBA-BU-BLAZAR program\footnote{http://www.bu.edu/blazars/VLBAproject.html}). 
The data are calibrated and imaged as presented in \cite{Jorstad2017}.\footnote{images and 
calibrated data of 3C~279 at all epochs can be found at www.bu.edu/blazars/VLBA\_GLAST/3c279.html.}
We have analysed the total and polarized intensity images during the period from 2007 June to 2018 August. This
results in 111 images in each Stokes parameter, $I$, $Q$, and $U$. The total intensity images were modelled by circular components 
with Gaussian brightness distributions in the same manner as described by \cite{Jorstad2017} using the routine \texttt{modelfit} in the \texttt{Difmap} 
software package \citep{Shepherd1997}. Each $I$ image is modelled by a number of components (knots), with parameters that result in the best fit to the $(u,v)$ data 
according to a $\chi^2$ test. Each knot is characterized by its
flux density $F$, distance from the core, $R$, position angle wrt the core, $\Theta$, and size (FWHM).
Uncertainties in the parameters are determined according to the formalism given in \cite{Jorstad2017}. 
The core, A0, is considered a stationary feature, located at the northeast (narrow) end of the jet structure. 

We have also modelled Stokes $Q$ and $U$ visibility data for components detected in the total intensity images,
following the technique described in \cite{Jorstad2007}. For comparison with
multi-wavelength light curves and 
polarization curves analysed in this paper, we have calculated at each epoch total and linearly polarized 
flux densities and electric vector position angles, EVPA, by summing the $I_i$, $Q_i$ and $U_i$ flux densities 
from all components $i$ of the source at each epoch. We have constructed a total 
flux density light curve at 43~GHz, 
$F_{\rm total}$, representing the sum of the flux densities of all components, and curves of the degree of polarization in percentage, $P=\sqrt{(F_{Q,{\rm total}}^2 + F_{U,{\rm total}}^2)}/F_{\rm total}\times 100$ percent, 
and EVPA in degrees, EVPA=$0.5{\rm arctan}(U_{\rm total}/Q_{\rm total})$. The light curve at 43~GHz
is presented in Figure~\ref{lc_radio}, and the polarization parameters at 43~GHz versus 
epoch are given in Fig.~\ref{optical_radio_polar}.  

\subsection{Optical polarimetry}
In this study, we include optical polarimetric data obtained with telescopes at the Crimean Astrophysical 
Observatory, St.~Petersburg University, Lowell Observatory (Perkins Telescope),
Steward Observatory, and Calar Alto Observatory.
The Galactic latitude of 3C~279 is 57\degr and $A_V$ = 0.078 mag, so that interstellar polarization (ISP) in
its direction is $<0.3$ per cent. To correct for ISP, the mean relative Stokes parameters of nearby stars were
subtracted from the relative Stokes parameters of the object. This removes the instrumental polarization as well, 
under the assumption that the stellar radiation is intrinsically unpolarized.
The fractional polarization has been corrected for statistical bias, according to \citet{Wardle1974}. For some of our analysis (see \S\ref{pol_behaviour}), we resolve the 
$\pm180\degr$ ambiguity of the polarization electric-vector position angle (EVPA) 
by adding/subtracting $180\degr$ each time that the subsequent value of the EVPA
is $>90\degr$ less/more than the preceding one. 

\subsection{Spectroscopic observations}
\label{sec:dct}
We performed spectral observations of 3C~279 with the 4.3~m Discovery Channel Telescope (DCT, Lowell Observatory, telescope located in Happy Jack, AZ, USA) using the DeVeny spectrograph\footnote{\url{https://jumar.lowell.edu/confluence/pages/viewpage.action?pageId=23234141\#DCTInstrumentationCurrent\&Future-DeVeny}} and Large Monolithic Imager (LMI)\footnote{\url{ https://jumar.lowell.edu/confluence/pages/viewpage.action?pageId=23234141\#DCTInstrumentationCurrent\&Future-LMI}}. The DeVeny spectrograph, with 300 grooves per mm and a grating tilt of 22\fdg35, was employed
to produce a spectrum from 3500\AA\, to 7000\AA, centred at 5000\AA, with a dispersion of 2.2~\AA/pixel and a pixel
size of 0.253 arsec. We used a slit width of 2.5 arcsec. 
The technique for the spectral observations and data reduction was adapted from that 
developed at the NASA Infrared 
Telescope Facility for observations with SpecX \citep{Vacca2003}, which is based on an observation of a `telluric standard'
whose intrinsic spectrum is known. This allowed derivation of a system throughput curve, which was then applied to the spectrum of the source. Each spectral observation of 3C~279 included 2-3 exposures of 900~s each. A calibration star, HD~112587, of spectral type A0 and located $<1\degr$ from 3C~279, was observed just before and after the quasar, with two exposures of 60~s each. The exposure lengths of the target and telluric standard varied slightly, depending on the brightness of 3C~279 and weather conditions. Since the DCT is capable of switching between different instruments within 2-3~minutes, the spectral observations were followed by photometric observations of the quasar in $R$ and $V$ bands with the LMI in order to determine the flux calibration. Bias and flat-field images were obtained for corresponding corrections for both instruments.


\section{RESULTS AND DISCUSSION}\label{results}

\subsection{Flux and colour evolution}\label{color_evolution}
\begin{figure}
\begin{center}
  \includegraphics[width=\columnwidth,clip]{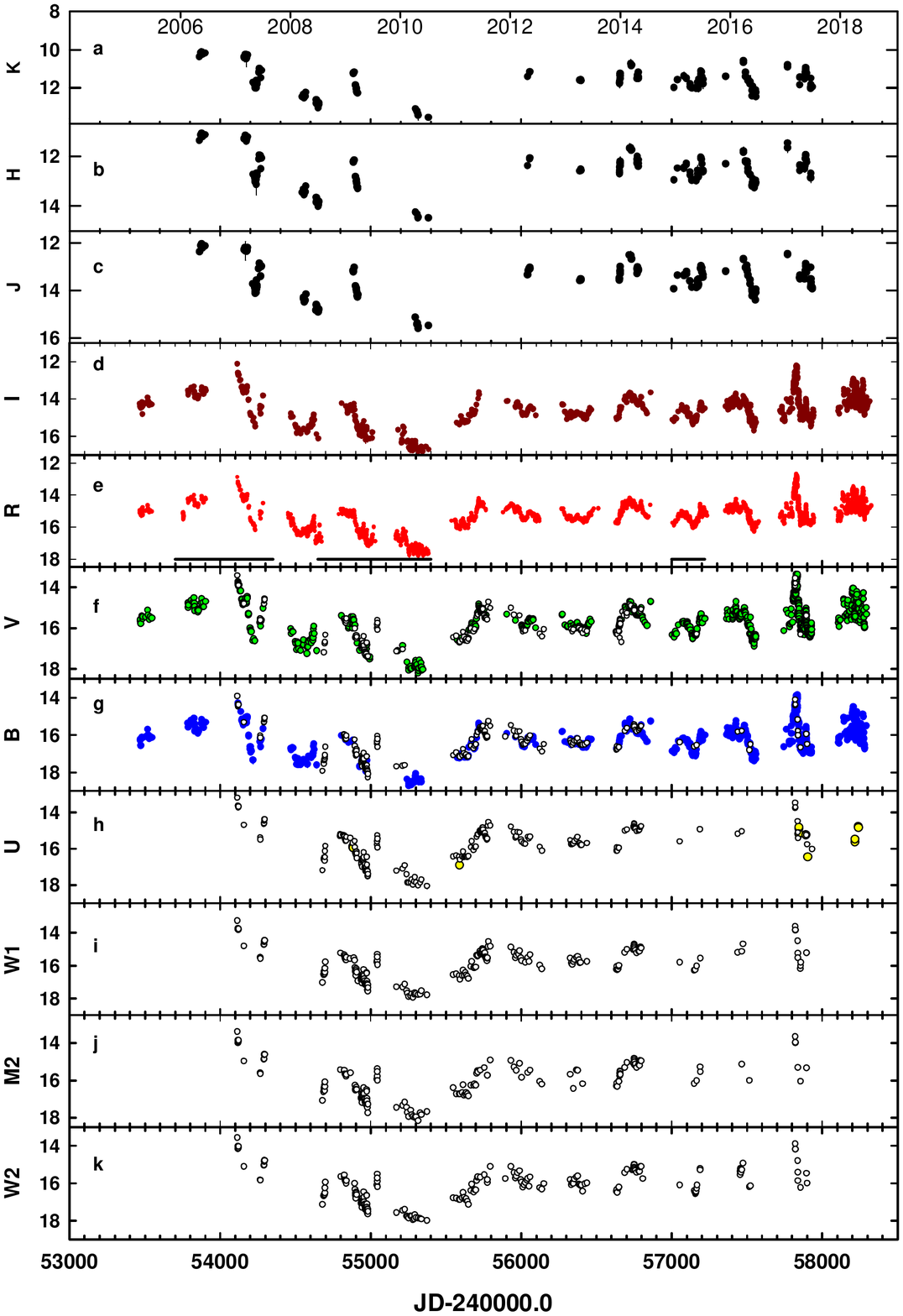}
   \caption{Optical, near-infrared, and ultraviolet light curves of 3C~279 over the time interval of the WEBT campaign. Horizontal lines in panel (\emph{e}) mark the data published in earlier WEBT-GASP papers \protect\citep{Bottcher2007, Larionov2008, Hayashida2012, Pittori2018}.
Empty circles in panels (\emph{f}) through (\emph{k}) refer to the \emph{Swift} data}.
\label{lc_UV_NIR}
\end{center} 
\end{figure}

The optical light curves of 3C~279, collected by the WEBT participating teams during the 2007--2018 time interval,
are shown in Fig.~\ref{lc_UV_NIR} (\emph{a-k}); horizontal lines in panel (\emph{e}) mark the data published in earlier 
WEBT-GASP papers \citep{Bottcher2007, Larionov2008, Hayashida2012, Pittori2018}. In panels (\emph{f-k}) of the same 
figure we show \emph{Swift} UVOT data (open circles). As is common for the WEBT campaigns, the data coverage in the
optical range is dense (nearly 5000 data points in $R$ band, 1000--1500 in $B$, $V$ and $I$) throughout each 
observational season. The most intensive observations were performed during high-activity states of the blazar.

\begin{figure}
\begin{center}
  \includegraphics[width=\columnwidth,clip]{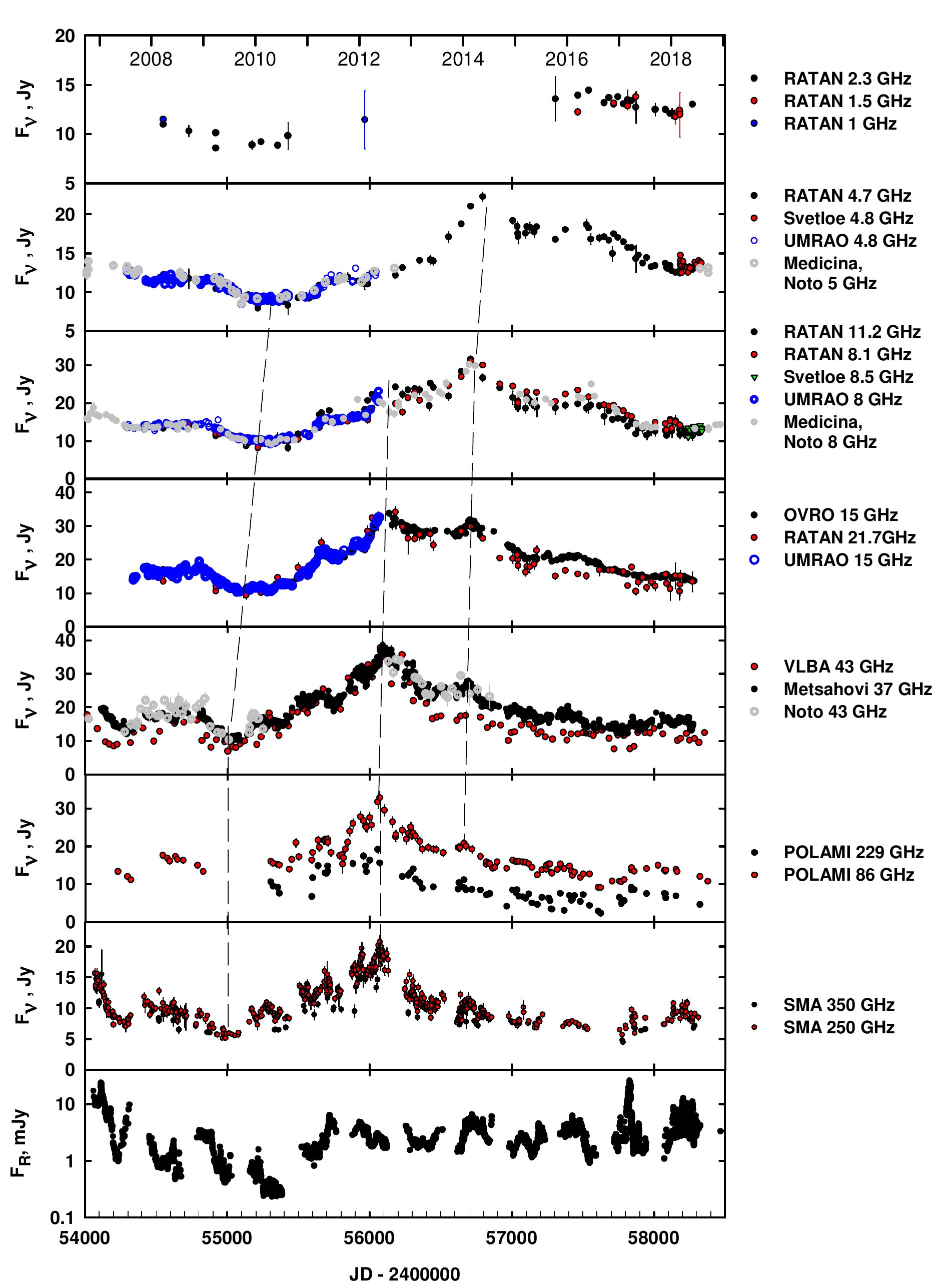}
   \caption{Radio light curves of 3C~279 over the time interval 2007--2018. For comparison, the $R$ band light curve is reproduced in the bottom panel. Dashed slanted lines connect positions of main extrema of the light curves (see \S~\protect\ref{radio_corr}).}
\label{lc_radio}
\end{center} 
\end{figure}

Radio flux densities versus time are plotted in Fig.~\ref{lc_radio}. The 43~GHz light curve is constructed based
on the VLBA images as described in \S \ref{vlba}. Slanted dashed lines connect minima 
and maxima in the different light curves, which may correspond to the same emission events 
of 3C~279 at different frequencies.

\begin{figure}
\begin{center}
  \includegraphics[width=\columnwidth,clip]{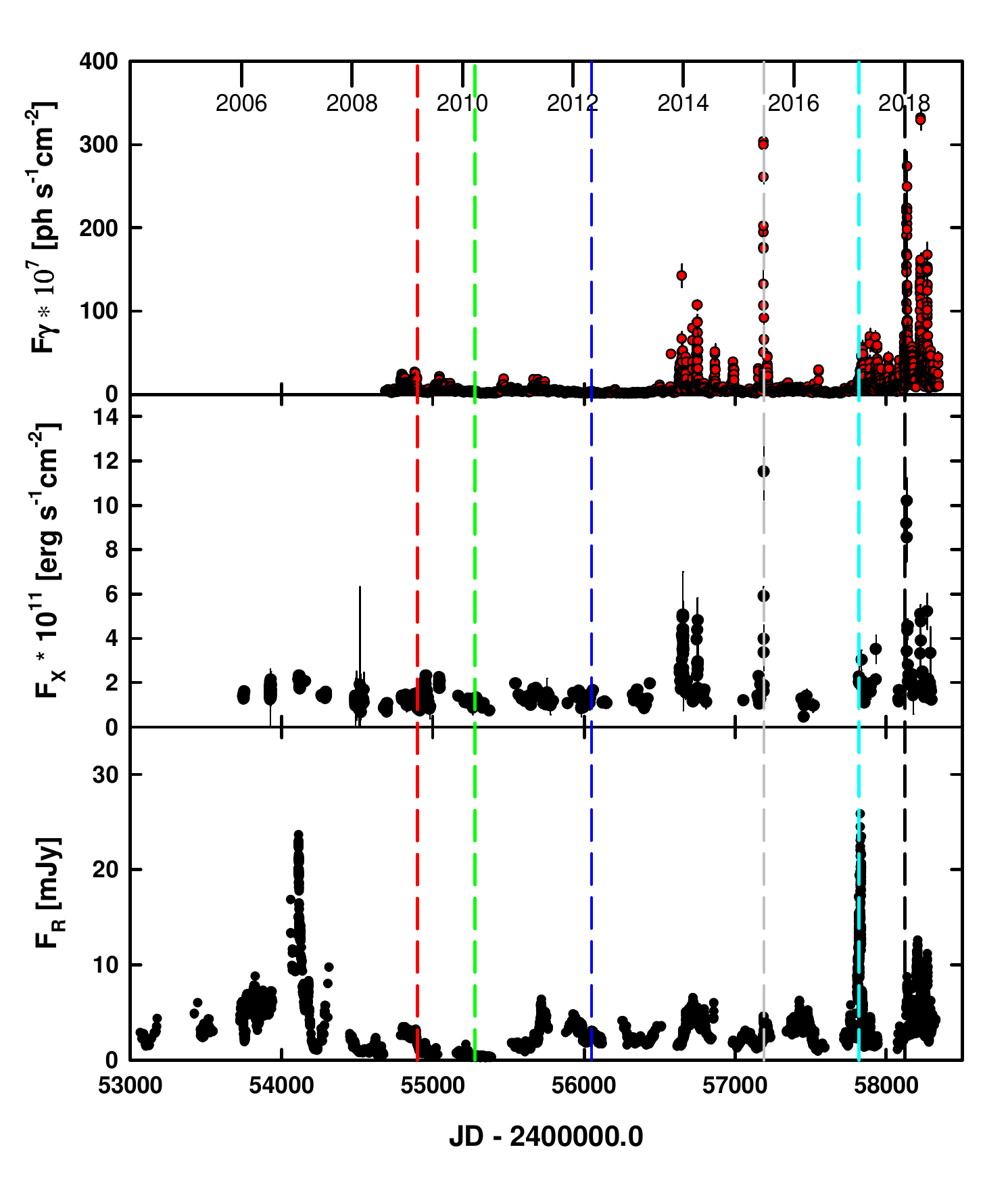}
   \caption{High-energy light curves of 3C~279 over the time interval 2006--2018. For comparison, the $R$ band light curve is reproduced in the bottom panel. Dashed vertical lines mark epochs when SEDs were constructed (see \S~\protect\ref{sect:sed}).}
\label{lc_gamma_X-ray}
\end{center} 
\end{figure}

We plot the high-energy (\emph{Fermi} LAT and \emph{Swift} XRT) light curves in Fig.~\ref{lc_gamma_X-ray}. 
Visual inspection reveals that the general patterns of the $\gamma$-ray and X-ray light curves are quite similar.

The question of whether a blazar's optical radiation becomes redder or bluer when it brightens is a topic of numerous papers.
It is commonly agreed that the relative contributions of the big blue bump (BBB) and Doppler-boosted synchrotron
radiation from the jet differ between quiescence and outbursts,
and that this is one factor that leads to variability of the SED. In a recent paper, \citet{Isler2017} have parametrized optical-NIR variablity of 3C~279 in terms of the combined contributions of the accretion disc and the jet. However, most previous studies are qualitative rather than quantitative, owing to the difficulty in evaluating the contributions of constant and slowly-varying components, such as starlight from the host galaxy, the accretion disc, and the broad emission-line region.

A straightforward way
to isolate the contribution of the component of radiation that is variable on the shortest time scales (presumably,
synchrotron radiation) has been suggested  by Hagen-Thorn \citep[see, e.g.,][and references therein]{Hagen-Thorn2008}.
 The method is based on plots of (quasi)simultaneous flux densities
in different colour bands and the construction of the relative continuum spectrum based on the slopes of the sets of flux-flux
relations thus obtained. This method was successfully applied for the quantitative analysis of the synchrotron radiation of several blazars, for example, 0235+164 \citep{Hagen-Thorn2008}; 3C~279, BL~Lac, and CTA~102 \citep{Larionov2008, Larionov2010,Larionov2016a}; 3C~454.3 \citep{Jorstad2010}; and 3C~66a, S4~0954+65, and BL~Lac \citep{Gaur2019}.

\begin{figure}
\begin{center}
  \includegraphics[width=\columnwidth,clip]{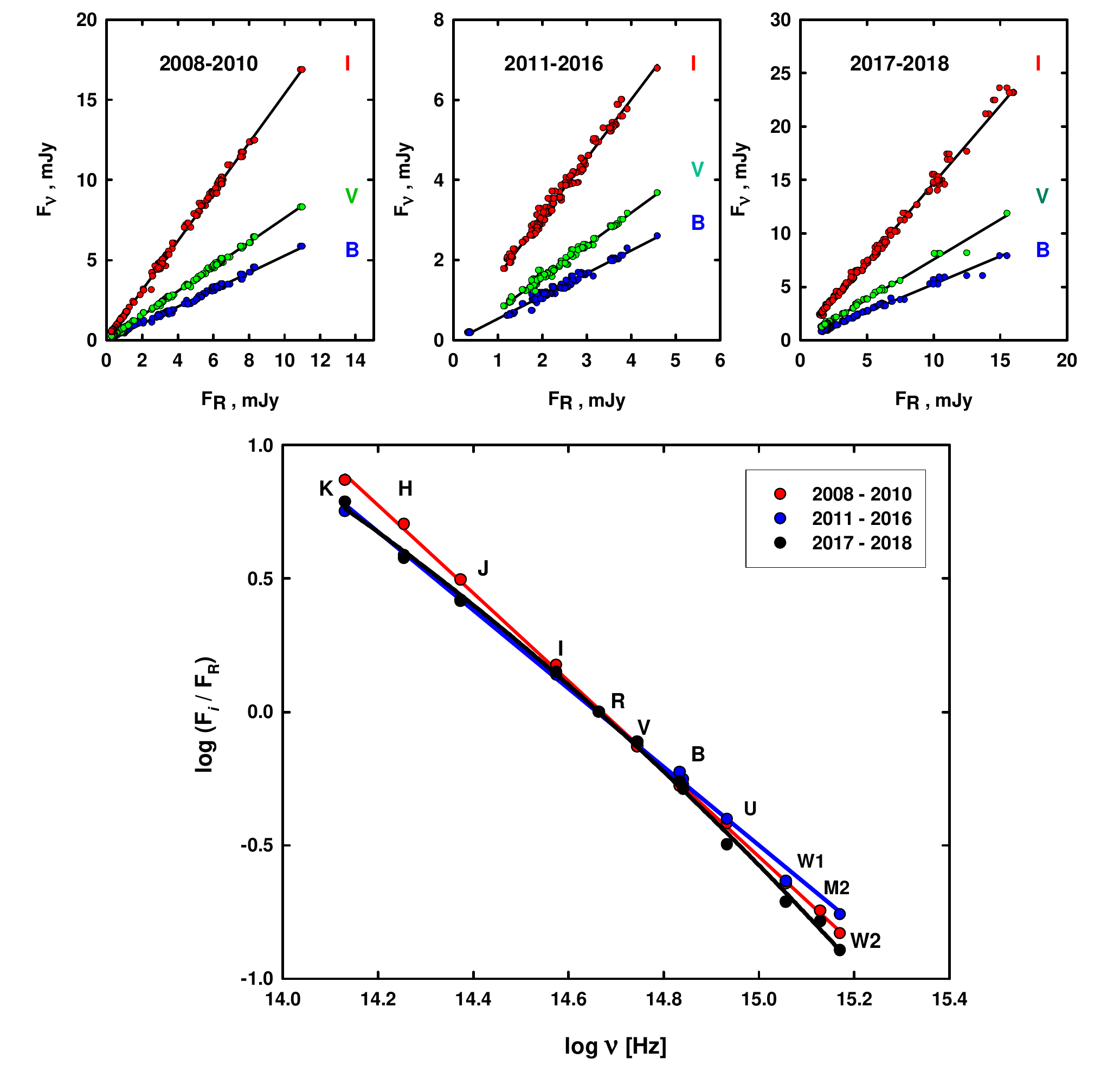}
   \caption{Flux-flux dependencies of $F_B$, $F_V$ and $F_I$ vs $F_R$ (upper row of panels). Relative continuum spectrum of variable component of radiation from UV through NIR bands (bottom panel).}
\label{SED_variable}
\end{center} 
\end{figure}

We adopt the same approach, as displayed in Fig.~\ref{SED_variable}, where the flux densities of 3C~279 in $B$, $V$ and $I$ bands are
plotted against the $R$-band flux density. The three plots in the top panels, obtained during different intervals of enhanced activity, together with analogous dependencies in UV and NIR bands, allow a derivation of the relative spectrum of the variable component, plotted in the bottom panel of the same figure. With this, we are able to trace the seasonal changes of the spectral index $\alpha$ (in the sense
$F_\nu \propto \nu^{-\alpha}$) over
the entire UV--NIR range.  We split the time range into three parts, determined via visual inspection of Fig.~\ref{lc_UV_NIR}: 2008--2010 (period of general decline of the flux, down to the unprecedented low level of 2010), 2011--2016 (modest level of activity over the optical range), and 2017-2018 (enhanced mean flux and short time-scale variations in
the optical bands). We obtain $\alpha= 1.65\pm0.02$ for 2008--2010, $\alpha= 1.47\pm0.02$ for 2011--2016, and $\alpha= 1.60\pm0.02$ for 2017--2018. The values of the slopes refer to the central frequency, corresponding to $R$ band. We emphasize that these values are for the \textit{variable} component only, not for the total flux density. We also find mild curvature (convexity) of the spectrum of the variable (synchrotron) component over the interval 2017--2018, i.e., a softening of the spectrum, despite elevated flux levels across the entire UV-optical frequency range (see Figures~\ref{lc_UV_NIR},~\ref{lc_gamma_X-ray}). We note that a very similar value of  $\alpha= 1.58\pm0.01$ in 2006--2007 was reported by \cite{Larionov2008}.

\subsection{Optical spectra and variability of broad emission-line clouds}
\label{spectra}

\begin{figure}
   \centering
  \includegraphics[width=\columnwidth,clip]{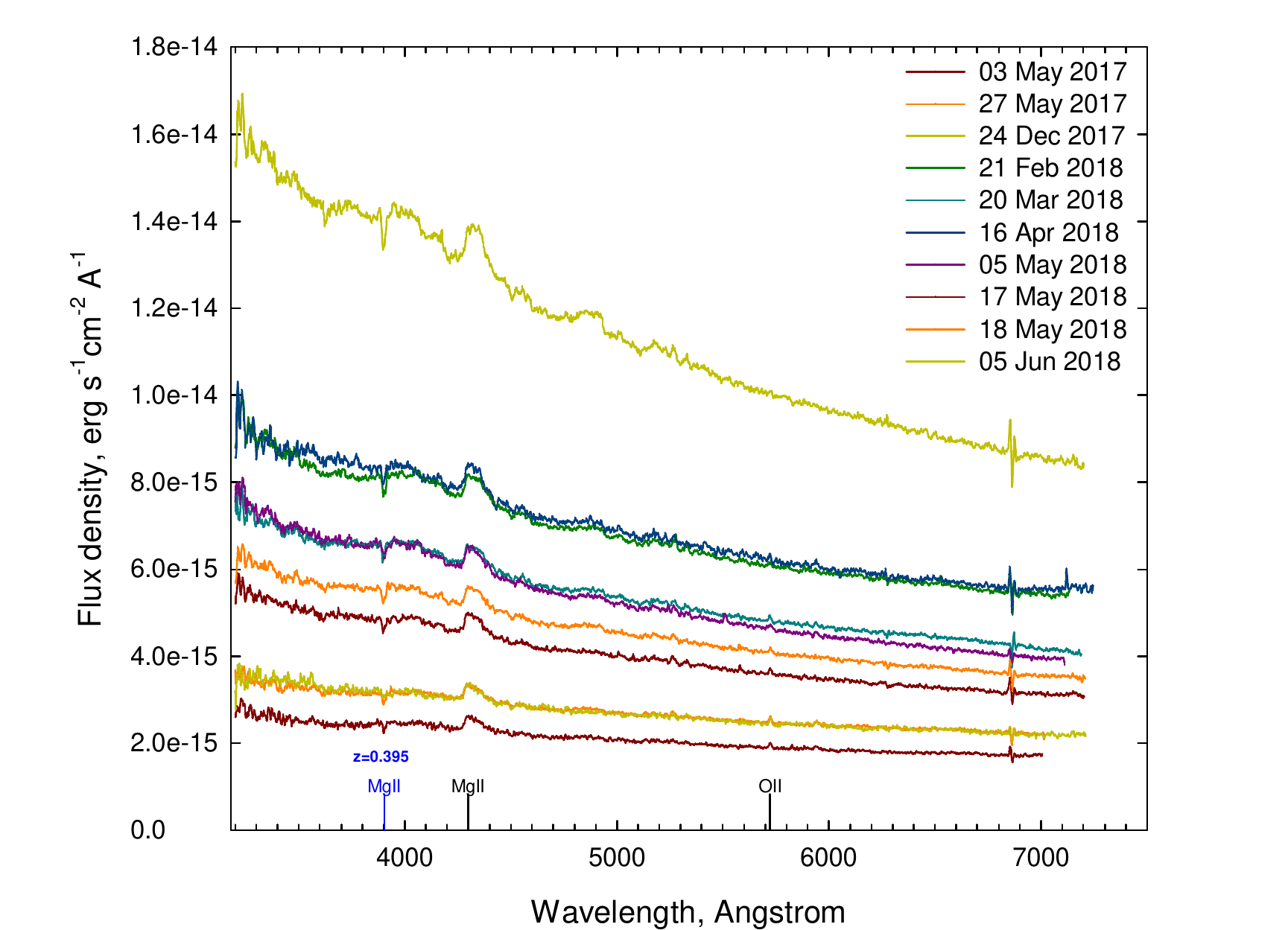}
      \caption{Spectra of 3C~279 during the 2017 and 2018 observing seasons.}
         \label{3c279spectra}
   \end{figure}

We analyse the optical spectroscopic behaviour of 3C~279 using data obtained at the 4.3~m 
Discovery Channel Telescope (DCT). Figure~\ref{3c279spectra} displays our spectra of 
3C~279 from the 2017 and 2018 observing seasons. All of these spectra contain a prominent 
\ion{Mg}{ii} $\lambda 2800$\AA\, broad emission line redshifted to $\lambda 4280$\AA. In addition to this prominent feature, we also mark in the figure the low-redshift absorption 
line of \ion{Mg}{ii} at $\lambda 3906$\AA\, arising in a foreground galaxy at $z=0.395$ \citep[see][]{Stocke1998}. We also mark an emission line of \ion{O}{ii}, intrinsic to 3C~279. We note that there are a number
of broad spectral features between $\lambda 3600$ and 5300{\AA} (rest wavelengths in the
2350-3450{\AA} range), often called the `little blue bump' and attributed to many
blended Fe lines \citep[e.g.,][]{Vestergaard2001}.

\begin{figure}
   \centering
  \includegraphics[width=\columnwidth,clip]{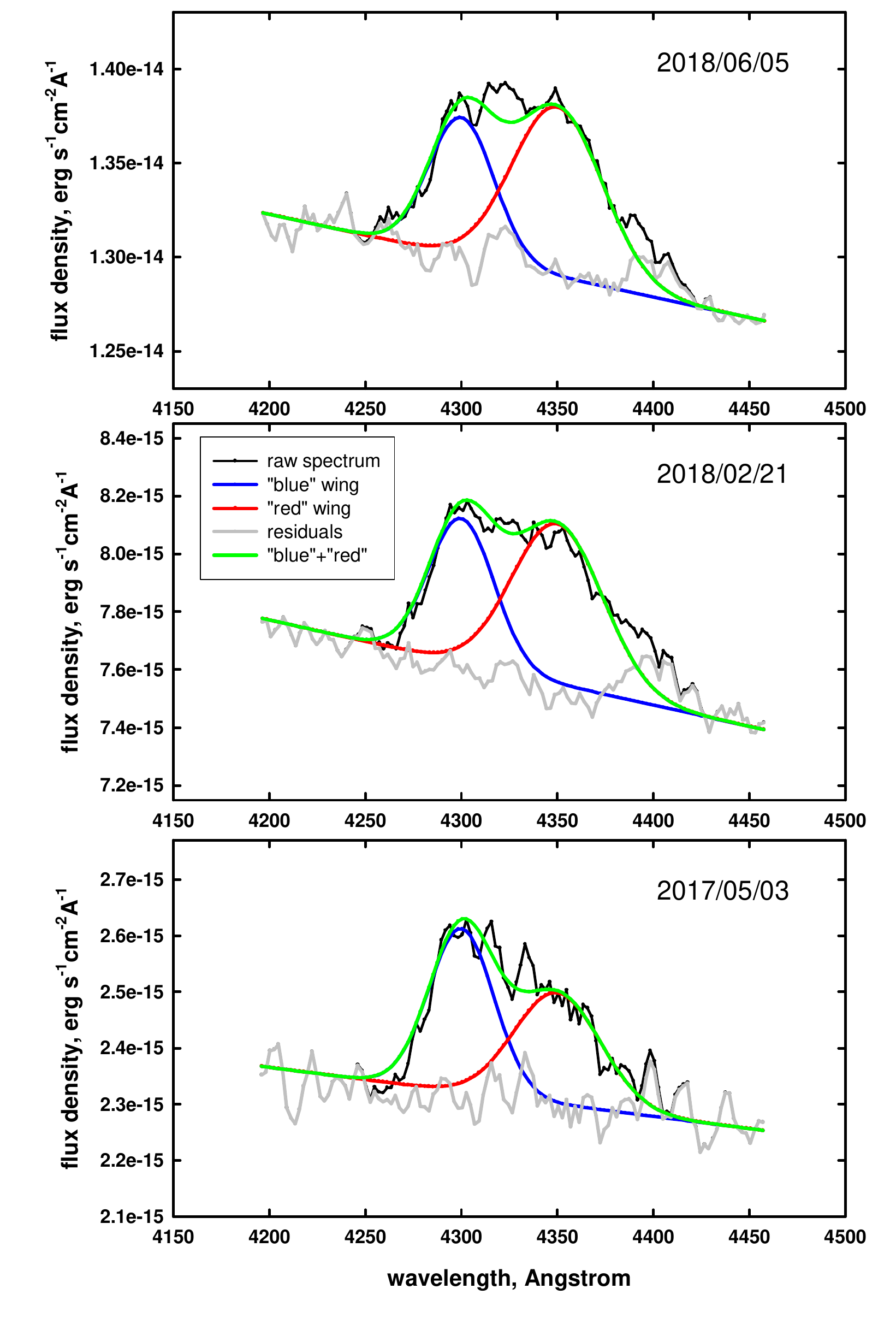}
      \caption{Examples of Gaussian fitting of the \ion{Mg}{ii} line profile for different levels of the continuum flux.}
         \label{MgIIgaussfit}
   \end{figure}

\begin{figure}
   \centering
  \includegraphics[width=\columnwidth,clip]{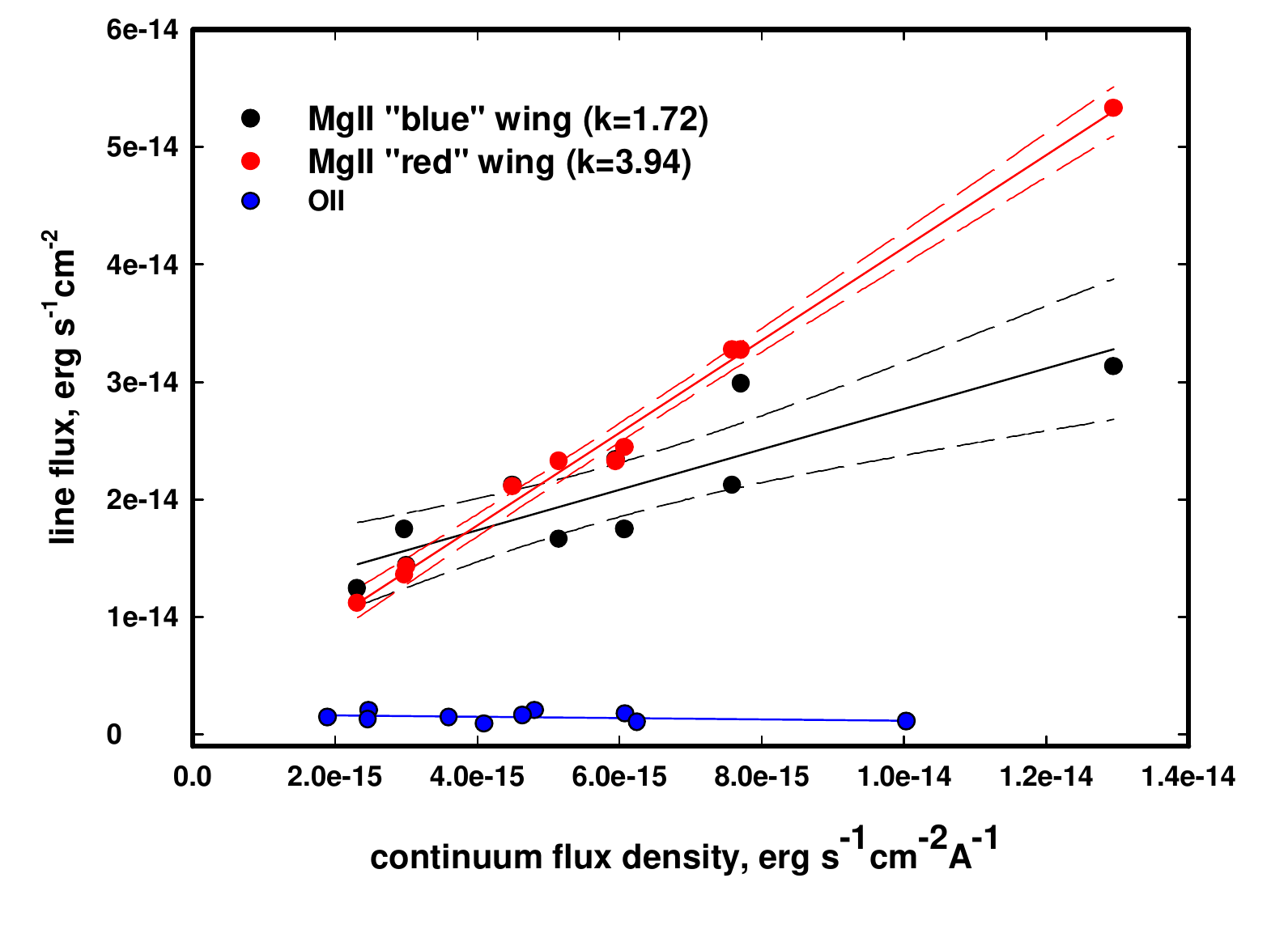}
      \caption{Dependencies of \ion{Mg}{ii} line fluxes vs.\ continuum flux density. The 95\% confidence intervals are also shown. For comparison, the dependence of [\ion{O}{ ii}] $\lambda_\mathrm{rest}$ 3727\AA\, flux vs.\ the adjacent continuum flux density is
      plotted near the bottom.}
         \label{MgIIblue_red}
   \end{figure}

Visual inspection of the spectra displayed in Figure \ref{3c279spectra} reveals that (1) the line flux of \ion{Mg}{ii} correlates with the continuum flux and (2) there is marked asymmetry (`red wing') to
this line, as previously noted by \citet{Punsly2013}. We deblend the observed line profiles, fitting them with two Gaussian functions superposed on a featureless continuum. Examples of the fit are given in Figure~\ref{MgIIgaussfit}. The wide range of continuum flux densities, seen in Figure~\ref{3c279spectra}, allows us to plot the dependencies of the fluxes in the `blue' and `red' components of \ion{Mg}{ii} on the continuum flux; see Figure~\ref{MgIIblue_red}. This definitive correlation is similar to that found in the  \ion{Mg}{ii} line of quasar 3C~454.3  by \citet{Leon-Tavares2013}. Less pronounced line flux variability in the spectra of several blazars has also been reported by \citet{Isler2015}. In contrast, stability of emission-line fluxes has been reported in several previous studies: 3C~454.3 \citep{Raiteri2008}, PKS~1222+216 \citep{Smith2011},  4C~38.41 \citep{Raiteri2012}, OJ~248 \citep{Carnerero2015}, and CTA~102 \citep{Larionov2016a}.


The two-component profile of the \ion{Mg}{ii} line in 3C~279 could be explained by radial 
infall of matter onto the accretion disc surrounding the central black hole. The speed of  the `red' component is then $\sim $3500 km~s$^{-1}$, corresponding to a $\sim$50~\AA\, wavelength shift of that component. The free-fall velocity of matter at a 
distance $R_{\rm pc}$ pc from a black hole of mass $M_9=10^9 $ M\sun\,  is
$v_{\rm in}= 2940 (M_9/R_{\rm pc}^{1/2})$ km~s$^{-1}$.
As reported in \citet{Nilsson2009}, the most reliable estimate of the black hole mass in 3C~279 is $10^{8.9}$ M\sun, hence the free-fall velocity is $\sim 3500$~km~s$^{-1}$ at
$R_{\rm pc}\sim 0.6$~pc.

Another approach \citep{Corbin1997} suggests that the redward profile asymmetries can be produced by the effect of gravitational redshift on the emission from a
`very broad line region,' provided that this region takes the form of a flattened ensemble of clouds viewed nearly face-on and with a mean distance of a few tens of gravitational radii from the black hole. 
The effect of gravitational redshift of the line emitted from a source at a distance $R_\mathrm{source}$ from the black hole is given by

\begin{equation}
\frac{\lambda_\mathrm{obs}}{\lambda_\mathrm{source}}= (1- \frac{R_\mathrm{Sch}}{R_\mathrm{source}})^{-1/2}
\label{eqn_grav}
\end{equation}
Here $R_\mathrm{Sch}=2GM/c^2$ is the Schwarzschild radius of the central black hole with mass $M$. 
The displacement of the red component of \ion{Mg}{ii} of $\sim$50\AA\, (see Fig.~\ref{MgIIgaussfit}) corresponds to a position of the emitting cloud of $\sim$43$R_\mathrm{Sch}\approx3\times10^{-3}$pc, or about 4 light-days, from the black 
hole. While this is closer than expected to the black hole for a cloud emitting 
\ion{Mg}{ii} lines,
which do not require high ionization parameters, we note that such small distances of
emission-line regions from
black holes have been inferred from microlensing studies of quasars \citep{Guerras2013}.
 
We measure the \ion{Mg}{ii} line FWHM for both the `blue' and `red' components, from which one can derive the velocity dispersion of the gas clouds in the broad-line region (BLR) 
and obtain, correspondingly, $v_\mathrm{blue}= 3450\pm200\:\mathrm{km\: s^{-1}}$ and $v_\mathrm{red}= 3800\pm200\:\mathrm{km\: s^{-1}}$ . These values are lower limits to the actual velocity range of the clouds, since the expected velocity dispersion depends on the geometry and orientation of the BLR~\citep[see, e.g.,][]{Wills1995}. In fact, because the line of sight to a blazar is probably nearly perpendicular to the accretion disc, the de-projected velocity range is likely to be a factor $\gtrsim 2$ higher than the FWHM given above if the BLR is in the 
disc.

The strong relationship between the optical (and presumably UV) continuum and the
emission-line flux suggests that the radiation responsible for the excitation of the broad 
\ion{Mg}{ii} lines comes mostly from the jet. This is difficult to reconcile
with the gravitational redshift hypothesis for the displacement of the red wing, since
the optical emission from the jet is unlikely to be highly beamed only several light-days
from the black hole \citep[see][]{Marscher2010}. If the dominant exciting radiation instead were to arise in the accretion disc, we would not expect to see such a strong correlation between the synchrotron continuum and line flux: the only direct association
of flare and superluminal knot production in the jet with an event in the accretion disc
corresponds to a \emph{decrease} in the optical-ultraviolet flux of the disc
(in the radio galaxy 3C~120; \citealt{Marscher2018}). Since the jet emission is expected to be
highly beamed by a Doppler factor $\sim 20$ \citep{Jorstad2017}, the association of
variations in emission-line flux with beamed synchrotron radiation from the jet implies 
that the  clouds responsible for these lines lie within $\sim 10\degr$ of the jet axis, 
as proposed by \citet{Leon-Tavares2015}.

The deblending of the \ion{Mg}{ii} line discussed above does not take into account 
possible contribution of time-variable \ion{Fe}{ii} emission lines. \ion{Fe}{ii} lines are 
present over a wide range of wavelengths near the \ion{Mg}{ii} line and should be present 
under the physical conditions that produce strong \ion{Mg}{ii} emission. We note that \citet{Patino-Alvarez2018} conclude that \ion{Fe}{ii} emission is negligible in the
vicinity of the \ion{Mg}{ii} line in the spectra of 3C~279 that they have studied.
Nevertheless, given the presence of a strongly time-variable red wing to the \ion{Mg}{ii} 
line, which is difficult to explain physically \citep[see above and][]{Punsly2013}, we 
plan to carry out a more thorough analysis of Mg and Fe line emission in a future study.

Based on the above considerations, we tentatively conclude that the variable, displaced 
(by 3500 km~s$^{-1}$)  component of the \ion{Mg}{ii} emission line arises from infalling 
clouds located $\sim0.6$~pc from the black hole and outside the jet, but within $\sim10\degr$ of the jet axis.

\subsection{Spectral energy distributions}\label{sect:sed}

\begin{figure}
   \centering
  \includegraphics[width=\columnwidth,clip]{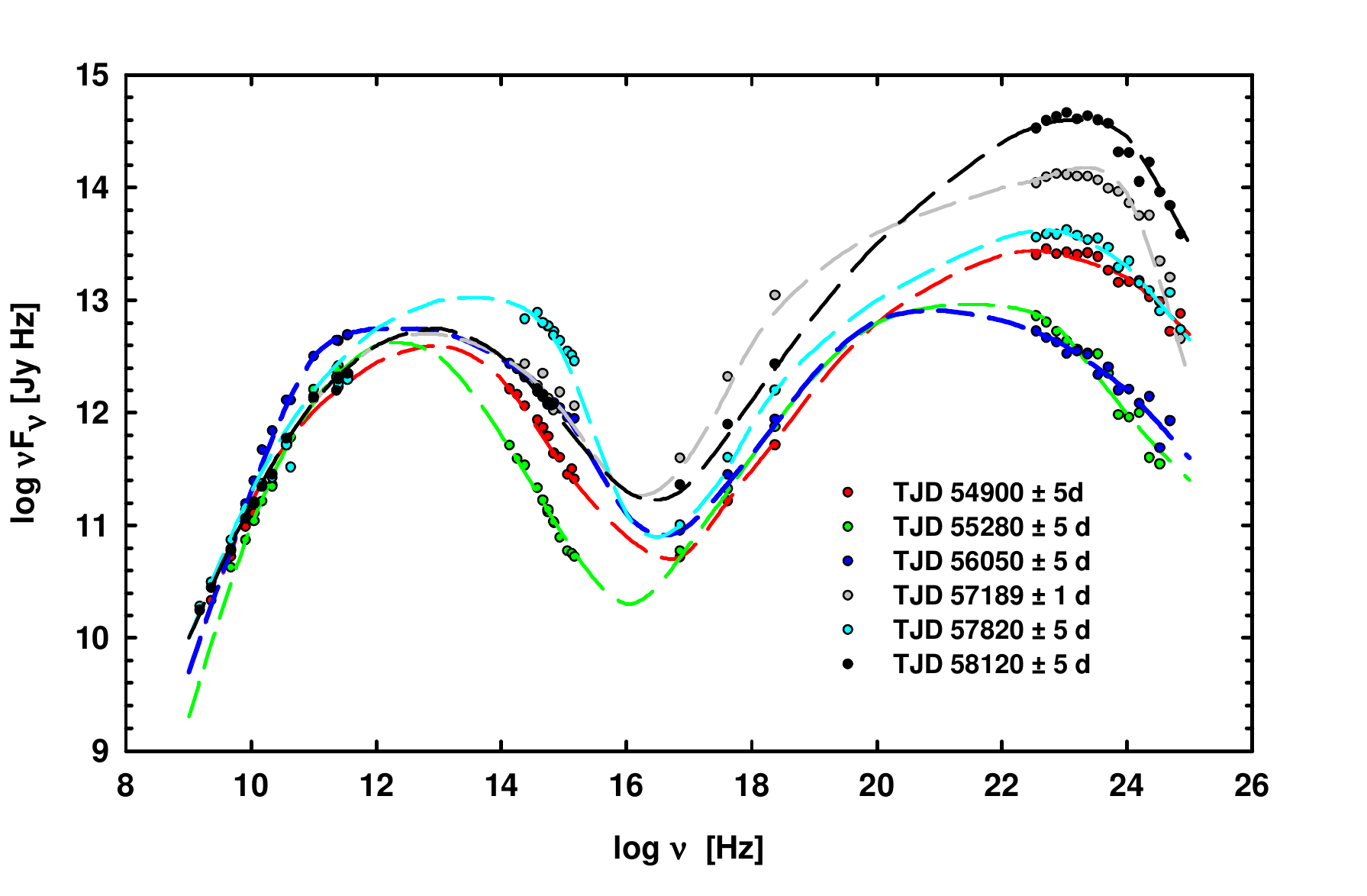}
      \caption{Quasi-simultaneous spectral energy distributions of 3C~279. TJD is
      truncated Julian date, JD$-$2400000}
         \label{3c279sed}
   \end{figure}

Figure \ref{3c279sed} presents the SED
(${\rm log}\,\nu F_\nu$ vs.\ {\rm log}\,$\nu$) of 3C~279 at 6 epochs with different levels 
of activity, displaying the usual double-hump shape. 
We follow the common interpretation that the humps correspond to synchrotron radiation at 
lower frequencies and inverse Compton (IC) scattering at high energies. The epochs, which 
are also marked in Figure
\ref{lc_gamma_X-ray}, include TJD~58120 when the $\gamma$-ray to optical-IR flux ratio at the corresponding peaks was $\sim60$, as well as other epochs when it was closer to unity. As has been discussed by \citet{Sikora2009}, the flux ratio should not greatly exceed unity if the $\gamma$-ray emission occurs via the
synchrotron self-Compton (SSC) mechanism corresponding to IC scattering of synchrotron photons from the jet, with the same population of electrons responsible for both processes. Otherwise, the seed photons for the scattering are likely generated
from outside the jet (external Compton, or EC, radiation). Among the six SEDs displayed in
Figure \ref{3c279sed}, only two epochs (TJD~57189 and 58120) are inconsistent with SSC 
$\gamma$-ray emission solely on this basis. One of these (TJD~57820) includes the highest 
optical flux found in our study; the flux of the peak of the $\gamma$-ray SED was only a 
factor of $\lesssim4$ higher than that of the synchrotron peak.

The SED should generally rise with frequency up to a point $\nu_{\rm peak}$ where the spectral index steepens to unity. This corresponds to the critical frequency of electrons
with energy per unit rest mass $\gamma_{\rm peak}$, at which the energy distribution 
steepens to $N(\gamma)\propto \gamma^{-3}$.
In some models, this occurs at a break in the injected electron energy 
distribution \citep[e.g.,][]{Sikora2009}, while in others there is a more 
gradual downturn of a log-parabolic energy distribution \citep[e.g.,][]{Massaro2006}.
In the model of \citet{Marscher2014}, the volume filling factor of the highest-energy
electrons is inversely proportional to energy owing to variations in magnetic field
direction relative to shock fronts. This causes a steepening of the energy distribution
when averaged over the entire volume. The shock model of \citet{Marscher1985} produces a 
break from radiative energy losses at an
energy that evolves with time as the shock moves down the jet.

If we tentatively assume that the $\gamma$-ray portion of the SED on TJD~57820 is produced 
by SSC emission, we can estimate the value of $\gamma_{\rm peak}$ as
\begin{equation}
\gamma_{\rm peak} \sim [\nu^{\rm SSC}_{\rm peak}/\nu^{\rm S}_{\rm peak}]^{1/2} \sim 4\times 10^4,
\label{eq:gamma}
\end{equation}
where we have taken $\nu^{\rm SSC}_{\rm peak}\sim 5\times10^{13}$ Hz (see Fig.\ 
\ref{3c279sed}). The corresponding magnetic field strength is
\begin{equation}
B \sim [\nu^{\rm S}_{\rm peak}(1+z)]/[(2.8\times 10^6~{\rm Hz})\gamma_{\rm peak}^2\delta]~{\rm G} \sim 4\times10^{-4}~{\rm G}
\label{eq:b}
\end{equation}
\citep[e.g.,][]{Rybicki1979}, where $\delta$ ($\sim40$ in 2017; see \S\ref{vlba} below)
is the Doppler beaming factor and $z=0.538$ is the redshift of 3C~279.
This is implausibly low for a compact emission region on parsec scales; hence we
conclude that even for the flare on TJD~57820 with such a high synchrotron amplitude,
the $\gamma$-ray emission was produced by the EC process. SSC emission could have
dominated on TJD~54900 and 55280 when the $\gamma$-ray to synchrotron flux ratio
was $\lesssim1$ and the value of $\nu^{\gamma}_{\rm peak}$ was relatively low, but
the absence of flux measurements at the SED peaks at these epochs precludes an
accurate assessment of this possibility.

In the case of EC emission, which we infer to apply to the outbursts of 3C~279
with the  highest fluxes \citep[as concluded earlier by][]{Sikora2009,Hayashida2015,
Ackermann2016}, the frequency of the peak in the $\gamma$-ray SED is given by
\begin{equation}
\nu^{\rm EC}_{\rm peak} \approx \nu'_{\rm seed} \gamma_{\rm peak}^2 \delta (1+z)^{-1} 
\propto \nu_{\rm seed} \Gamma \delta (1+z)^{-1},
\label{ECfreqpeak}
\end{equation}
where $\nu_{\rm seed}\approx \nu'_{\rm seed}\Gamma^{-1}$ is the frequency of the peak in 
the SED of the 
seed photons in the host galaxy rest frame, and $\Gamma$ is the bulk Lorentz factor
of the emitting plasma. Note that $\nu^{\rm EC}_{\rm peak}$ is approximately proportional 
to $\delta^2$ if $\Gamma\sim \delta$, as is expected to be the case for blazars. 
An increase in the Doppler factor should increase both the flux and frequency
of the SED peak. While an increase in the latter is not
evident in the SEDs displayed in Figure \ref{3c279sed}, the SEDs of very sharp flares
on time scales of minutes to hours in 3C~279 have exhibited an increase in
$\nu^{\rm EC}_{\rm peak}$ \citep{Hayashida2015,Ackermann2016} during maximum flux.

In general, a rise in flux involves some combination of an increase 
in the number of radiating particles, the magnetic field strength, the Doppler
beaming factor, and the energy density of seed photons, \textbf{$u_{\rm seed}$}.
Radiative energy losses
of the highest-energy electrons tend to be severe during outbursts in 3C~279
\citep{Sikora2009,Hayashida2015,Ackermann2016}, so that essentially all of the energy that 
the electrons obtain from particle acceleration is transferred to electromagnetic radiation. 
Increases in apparent luminosity require an increase in either the number of these electrons or the Doppler beaming factor.

Although the peak \textbf{frequency} of the synchrotron SED is determined only within about an order
of magnitude, this is sufficient to specify that the ratio of the IC to synchrotron
peak frequencies is $10^{9\pm0.5}$. The Lorentz factor (energy in rest-mass
units) of the electrons radiating at the EC and synchrotron SED peaks should be the
same and given by
\begin{equation}
\gamma_{\rm peak} \sim [\nu^{\rm EC}_{\rm peak}(1+z)/(\delta\nu'_{\rm seed})]^{1/2}
\label{gampeakEC},
\end{equation}
where $\nu'_{\rm seed}$ is the frequency of the maximum of the seed photon SED as
measured in the frame of the radiating plasma. Equation \ref{eq:b}
then allows us to derive the magnetic field strength.
Based on the TJD 57189 SED, when the peak IC flux was $\sim40$ times the peak synchrotron
flux, indicating dominance by the EC process, we express the parameters as
$\nu^{\rm S}_{\rm peak}= 10^{14}\nu^S_{\rm peak,14}$~Hz,
$\nu^{\rm EC}_{\rm peak}= 10^{23}\nu^C_{\rm peak,23}$~Hz,
$\nu'_{\rm seed}= 10^{16}\nu'_{\rm seed,16}$~Hz, and,
from apparent superluminal motions, $\delta= 20\delta_{20}$ and $\Gamma= 20\Gamma_{20}$ \citep{Jorstad2017}. Equation \ref{gampeakEC} then becomes
\begin{equation}
\gamma_{\rm peak}\sim900 [\nu^\gamma_{\rm peak,23}(1+z)/(\delta_{20}\nu'_{\rm seed,16})]^{1/2}.
\label{gampeakvalues}
\end{equation}

If the seed photons are primarily from the little blue bump \citep[the Ly$\alpha$ emission
line is not very strong in 3C~279;][]{Stocke1998}, so that $\nu_{\rm seed}\sim10^{15}$ 
Hz, $\nu'_{\rm seed,16}\sim 2\Gamma_{20}$,
$\gamma_{\rm peak} \sim 800$, and $B\sim 4$~G. Even with such a low value of
$\gamma_{\rm peak}$, the radiative cooling time of the electrons in the plasma frame is 
only $7.8\times10^8[B^2+[(u'_{\rm seed})^2/(8\pi)](\gamma'_{\rm peak})^{-1} \sim 40$~s
(dominated by EC losses that are $\sim 40$ times stronger than 
synchrotron losses given the flux ratios of the two SED humps), allowing for rapid 
variability (although limited by the light-crossing time across the region). If, on the 
other hand, the seed photons correspond
to blackbody radiation from dust with a temperature $\sim 1200$~K 
\citep[as in 4C21.35;][]{Malmrose2011}, $\nu'_{\rm seed,16}\sim 0.05\Gamma_{20}$,
$\gamma_{\rm peak} \sim 5000$, and
$B\sim 0.1$~G. In this case, the radiative cooling time is $\sim 10^4$~s in the plasma
frame and $\sim 10^3$~s in the observer's frame. Intra-day flux variations are therefore
possible in either case if the size of the emission region is
$\lesssim0.01\delta_{\rm 20}$ pc.

Another possible source of seed photons is synchrotron radiation from a relatively slow 
(and, correspondingly, less beamed) Mach disc \citep{Marscher2014} or sheath of the jet 
\citep{MacDonald2015}. As suggested by \citet{Marscher2010}, such slowly moving plasma 
could produce the requisite number of seed photons while being too poorly beamed to 
contribute substantially to the observed optical flux. Stacked multi-epoch VLBA images 
have confirmed the existence of sheaths in the jets of several blazars, including 3C~279 
\citep{MacDonald2017}. The peak of the SED of such synchrotron 
photons is likely to be in the far-IR range, in which case the above estimate 
of the magnetic field for seed photons from hot dust should be applicable.

A major difference between EC from polar emission-line clouds or slowly moving jet plasma and EC from 
dust is the possibility of variations in the former case, while dust is expected to provide 
a steady source of seed photons. Variability of the seed photons can explain the general
absence of repeated episodes of variability with nearly identical patterns of temporal
behaviour.
   
   \subsection{Inter-band correlations}
\label{correlations}

\subsubsection{$\gamma$-ray -- X-ray correlations}

\begin{figure}
   \centering
  \includegraphics[width=\columnwidth,clip]{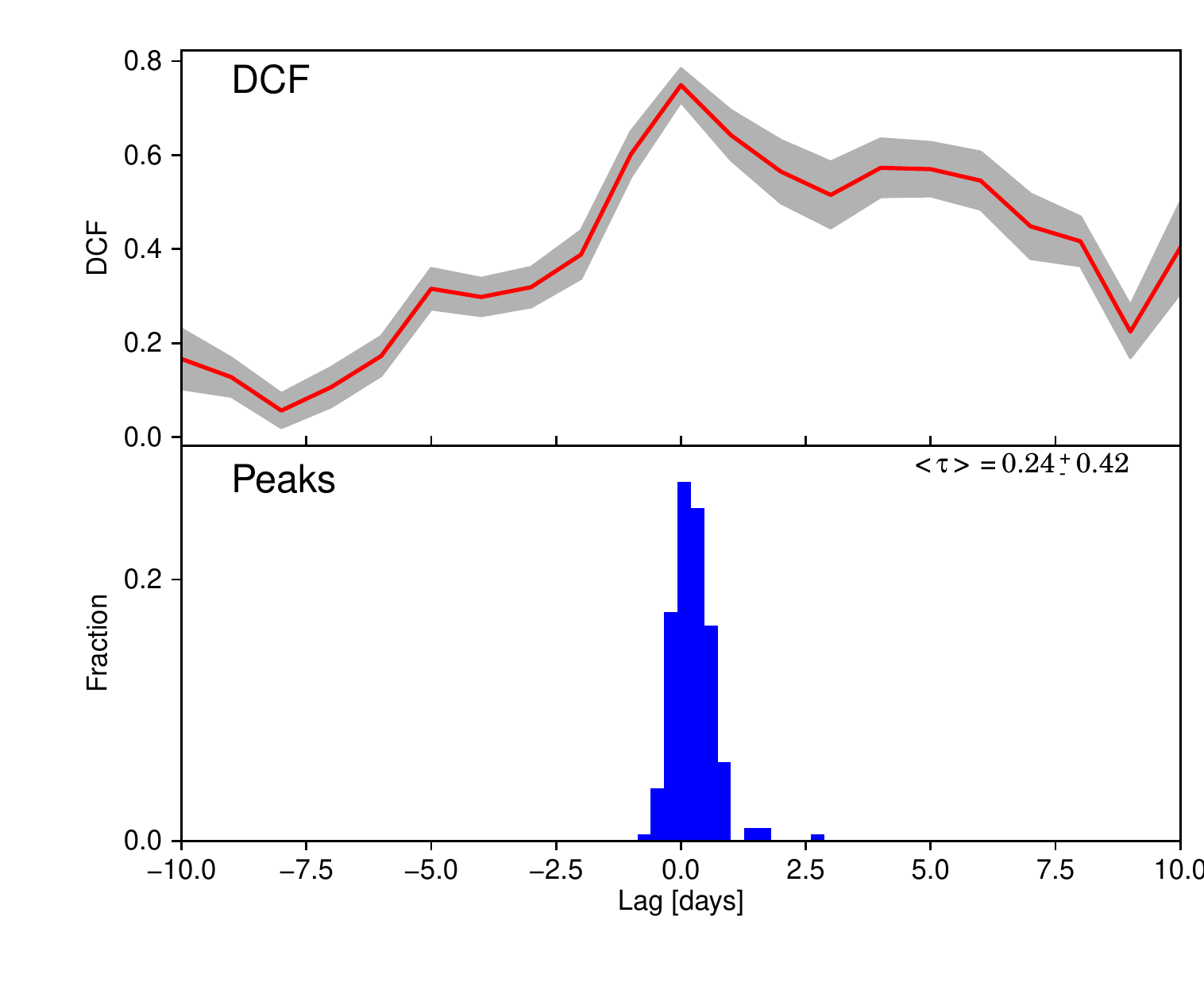}
      \caption{Upper panel: DCF between  $\gamma$-ray and X-ray light curves of 3C~279 in 2008--2018. Positive delays correspond to X-ray leading $\gamma$-ray variations.  Bottom panel: Distribution of lags of maxima of the DCF from 200 Monte Carlo simulations with flux redistribution and 67\% bootstrapping. The grey area in the upper panel correspond to $\pm 1 \sigma$ spread of simulated DCFs. }
         \label{dcf_gamma-X}
   \end{figure}

We calculate the discrete correlation function (DCF) \citep{Edelson1988, Hufnagel1992} between the
$\gamma$-ray and X-ray flux variations of 3C~279 during 2008--2018. We determine the significance of the correlation via the flux redistribution -- random sub-sample selection method \citep[see, e.g.][]{Peterson1998}.  The results are presented in Figure \ref{dcf_gamma-X}, which reveals a strong
correlation with a peak DCF of $0.75 \pm 0.03$ and a delay of the $\gamma$-ray behind X-ray variations of 0.24$\pm$ 0.42 days, statistically consistent with zero lag. This implies that the correlation is significant at a level higher than p=0.001 per cent \citep[e.g.,][]{Bowker1972} and conforms with Figure~\ref{lc_gamma_X-ray}, which shows that every X-ray flare has a $\gamma$-ray counterpart and \textit{vice versa}.

\subsubsection{$\gamma$-ray -- optical correlations}
\label{gamma-optical}

\begin{figure}
   \centering
  \includegraphics[width=\columnwidth,clip]{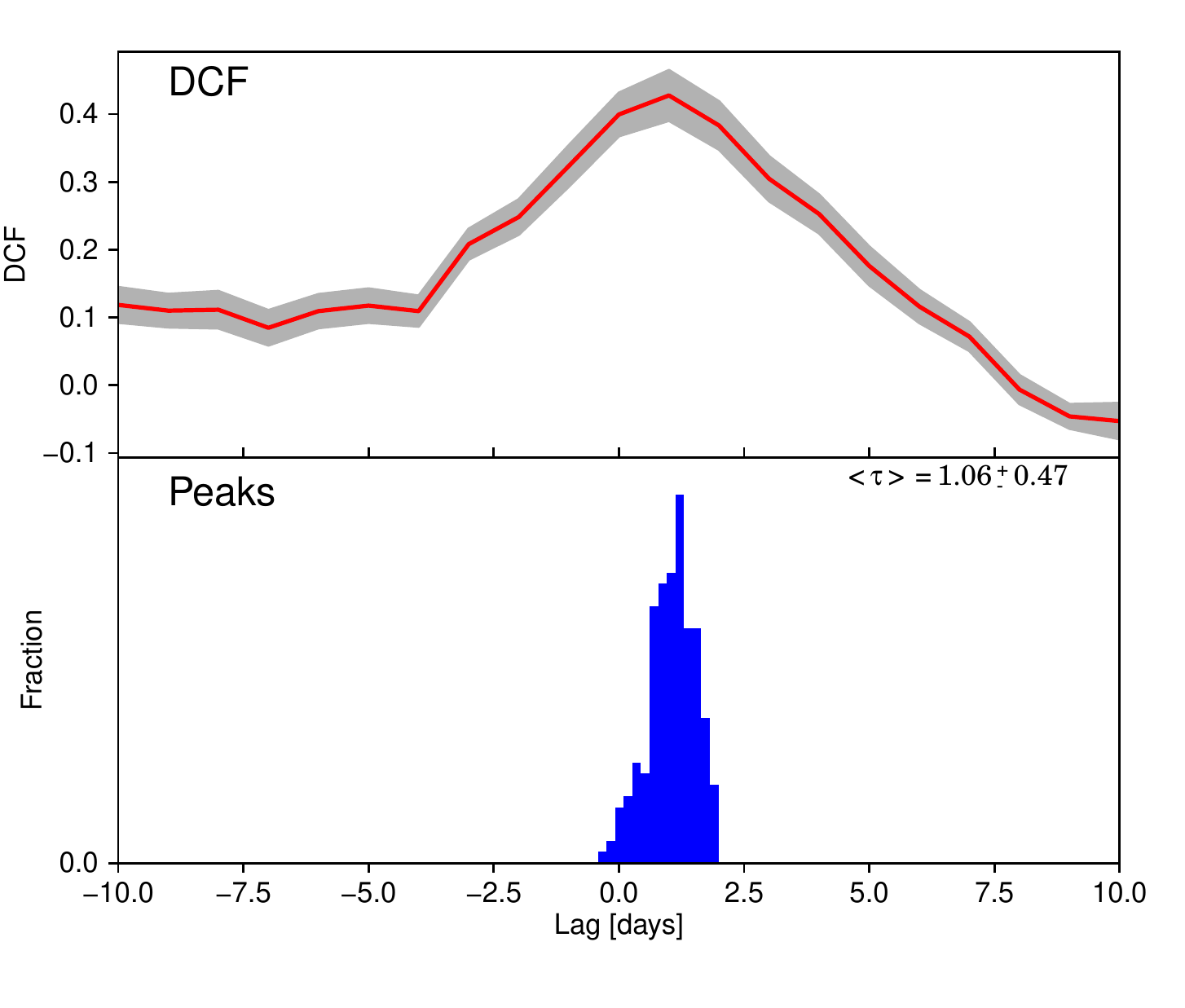}
      \caption{Upper panel: DCF between optical and $\gamma$-ray light curves of 3C~279 in 2008--2018. 
      Positive delays correspond to  $\gamma$-ray leading optical variations. 
      Bottom panel: Distribution of lags of maxima of the DCF after 200 Monte Carlo simulations with 
      flux redistribution and 67\% bootstrapping. The grey area in the upper panel correspond to $\pm 1 \sigma$ spread of simulated DCFs. }
         \label{dcf_R-gamma}
   \end{figure}

In a similar way, we calculate the DCF between the optical and $\gamma$-ray flux variations. The resulting correlation, given in Fig.~\ref{dcf_R-gamma}, is rather weak,
with a maximum DCF of 0.42. There is a delay on the level of $2\sigma$ between the variations in the two energy bands, with the $\gamma$-ray leading the optical variability by $1.06\pm 0.47$ days. The optical and $\gamma$-ray light curves of 3C~279 are quite complex, with `sterile'  (optical without $\gamma$-ray counterpart) and `orphan' ($\gamma$-ray without optical counterpart) flares occurring in both energy ranges.  To estimate the statistical significance of the correlation between $\gamma$-ray and optical variations, we have performed a Monte-Carlo simulation in the following way. At first, we approximated both $\gamma$-ray and optical light curves with a set of double-exponential functions in a manner similar to \citet{Abdo2010}. After obtaining the statistical distributions of peak parameters $(f_{\mathrm{max}}, t_{\mathrm{raise}}, t_{\mathrm{decay}})$, we generated a set of synthetic light curves by co-adding random peaks drawn from these distributions and placed into random positions. Any correlation between these synthetic light curves is by random chance, so by computing the correlation between them we can estimate the probability of a spurious correlation at a given level. Among $10^4$ artificial light curves generated in this manner, none gave a correlation coefficient equal to or higher than the observed value of 0.4. Therefore, we can consider this correlation level as statistically significant at the 0.01 per cent level.

\begin{figure}
\begin{center}
   \includegraphics[width=\columnwidth,clip]{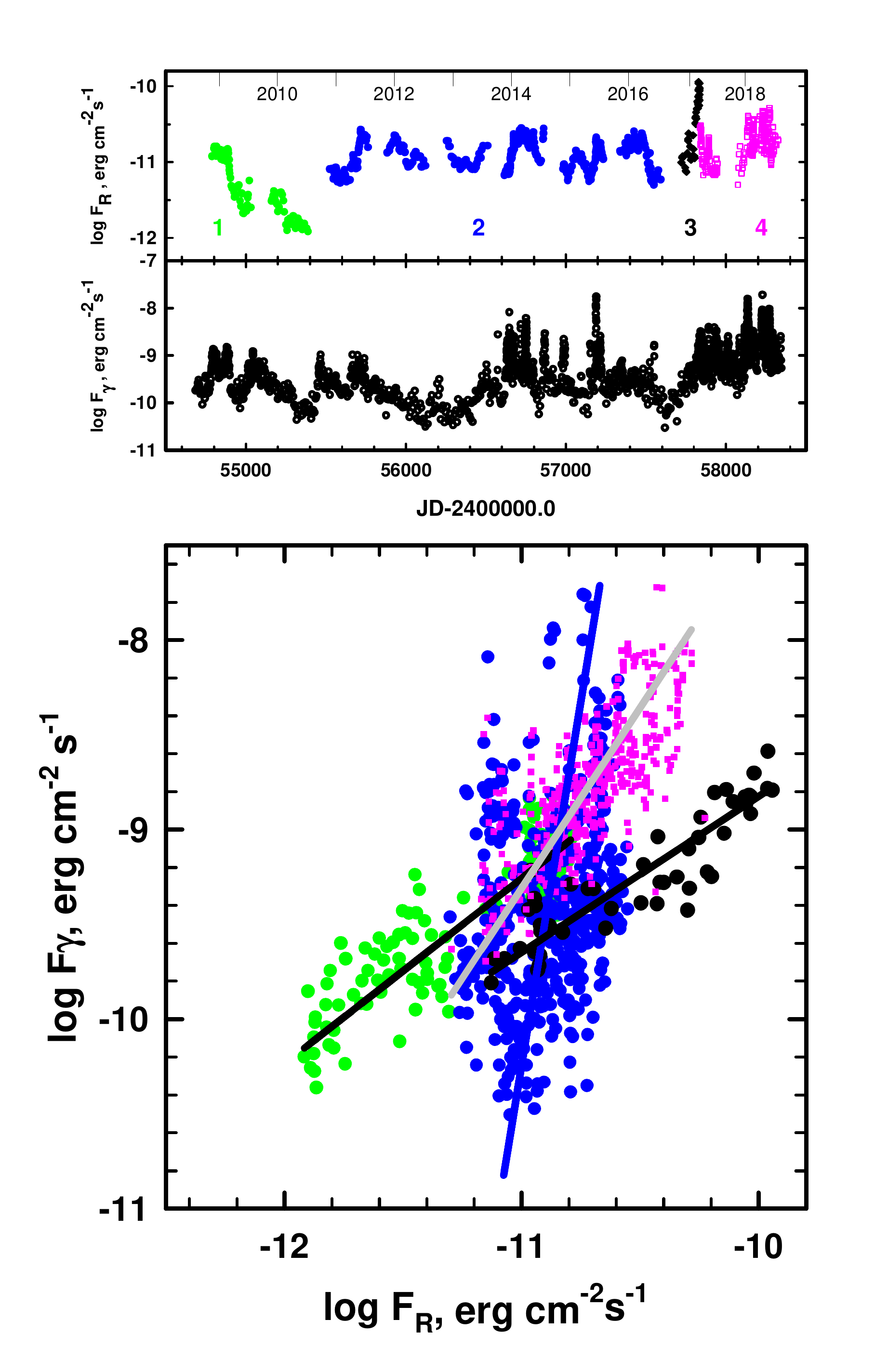}
        \caption{From top to bottom: $R$ band log flux density light curve, with different colours and numerical marks 
        corresponding to different stages of activity;  logarithmic $\gamma$-ray flux light curve; optical -- $\gamma$-ray flux-flux diagram. 
        Slopes of the linear (on a logarithmic scale) fits
        are (1) $0.97\pm0.04$,  (2) $7.7\pm1.2$, (3) $0.82\pm0.06$ and (4) $1.90\pm0.09$.}
\label{opt_gamma}
\end{center} 
\end{figure}

This short delay between $R$-band and $\gamma$-ray variations allows us to compare directly the optical and $\gamma$-ray light curves. To do this, we
bin the $R$-band optical data so that the mid-point and size (in time) of each optical bin corresponds to those of the respective $\gamma$-ray bin. Figure~\ref{opt_gamma}, where we plot the optical (upper panel) and $\gamma$-ray (middle panel) light curves, and $\gamma$-ray flux  versus $R$ band flux (bottom panel),
demonstrates clear differences in the optical/$\gamma$-ray relationship during the various stages of activity of 3C~279. The slopes of $\log F_\gamma - \log F_R$ dependencies, for time intervals (1) through (4) (as marked in the upper panel of that figure), are correspondingly: (1) $0.97\pm0.04$,  (2) $7.7\pm1.2$, (3) $0.82\pm0.06$ and (4) $1.90\pm0.09$. The slope of the dependence therefore changes dramatically, starting from  $\sim1$ (2008--2010), then switching to $\sim8$ (2011--2016). The onset of optical and $\gamma$-ray activity that occurred in late 2016 altered the slope again to $\sim1$. Shortly after the maximum of the optical outburst, the slope changed again, to $\sim2$. Remarkably, unlike transitions from the slope of 1 to 8 and back to 1, where the changes seem to have happened during seasonal gaps, the last change appears to have occurred over a few days, TJD~57837$\pm2$.  It is tempting to find some event(s) that accompanied this switch from linear to quadratic slope. The only change in other observed parameters that is simultaneous with this transition is a short-term drop of the optical polarization degree to 1.5\% during the night of optical maximum. However, as can be seen from Fig.~\ref{optical_radio_polar}, there are several other incidents of low polarization, with no obvious physical reason to connect these two events.

The $\gamma$-ray flux in Figure~\ref{opt_gamma} is derived by multiplying the
photon flux (see \S~\ref{sec:LAT}) by an integral over a fixed log-parabolic energy
distribution. It is therefore directly proportional to the photon flux, which in
turn is proportional to the flux density divided by frequency. 
The synchrotron flux is the flux density times the fixed
observed R-band frequency (which is above the frequency of the peak in the synchrotron
SED). From \emph{relativistic beaming alone}, we then expect the dependencies on Doppler 
factor to be $F_\gamma\propto \delta^{2+\alpha_{\gamma}}$ and
$F_{\rm opt}\propto \delta^{3+\alpha_{\rm opt}}$ \citep[see][]{Dermer1995}.
Since the peak of the $\gamma$-ray SED is usually in the LAT range except during
low states (cf. Fig.\ \ref{3c279sed}), we adopt $\alpha_\gamma=1.0$ for the $\gamma$-ray
spectral index, while in \S\ref{color_evolution} we derived the spectral index of the
variable optical component to be $\alpha_{\rm opt}=1.56\pm0.11$. 

As found by \citet{Sikora2009}, \citet{Hayashida2015}, and \citet{Ackermann2016}, the 
$\gamma$-ray and optical
emission is so luminous that the electrons lose nearly all of their energy to EC radiation 
on time scales shorter than the light-travel time across the emitting region. In this 
fast-cooling case, the ratio of $\gamma$-ray to optical flux (as defined above) is 
proportional to the ratio of EC radiative losses to synchrotron losses, which in turn is 
proportional to $u'_{\rm seed}/B^2 \propto u_{\rm seed}\Gamma^2/B^2$.
An increase in the number of radiating electrons $N_{\rm re}$ increases proportionately 
the luminosity of the dominant emission process, which is EC $\gamma$-ray production 
during the higher-flux states in 3C~279. The ratio of EC to synchrotron luminosity is affected by changes in the magnetic field $B$ or energy density of seed
photons (as measured in the emitting plasma frame), which can occur if (i) the plasma 
moves toward or away from the source of the seed photons, (ii) more seed photons are 
produced near the jet, or (iii) the bulk Lorentz factor $\Gamma$ changes (since the energy 
density of the seed photons in the plasma frame $u'_{\rm seed} \propto \Gamma^2$).

We can consider a number of cases of different physical parameters changing, each with
its own dependence between $\gamma$-ray and optical flux. We parametrize the optical/$\gamma$-ray flux relationship as $F_\gamma \propto F_{\rm opt}^{\zeta}$. Based
on arguments given in \S\ref{sect:sed}, here we assume that the EC process dominates the 
$\gamma$-ray production.
\begin{enumerate}
\item{} An increase solely in the number of radiating electrons $N_{\rm re}$ should cause the synchrotron and EC flux to rise by the same factor, so that $\zeta\approx 1$. The
frequencies of the SED peaks should remain the same unless there is a change in the energy 
at which the slope of the electron energy distribution becomes steeper than $-3$.
\item{} An outburst can result from a higher Doppler factor owing to bending toward the
line of sight of the emitting region. This can occur if the entire jet changes its direction (wobbles or precesses), or if different parts of the jet cross-section with 
various velocity vectors relative to the mean become periodically or sporadically bright 
as time passes. A possible realisation of such behaviour in the case of CTA~102 was 
suggested by \citet{Larionov2017}. The flux during such events should 
follow the beaming-only relationships given above, with $F_\gamma\propto \delta^3$ and
$F_{\rm opt}\propto \delta^{4.56}$, hence $\zeta \approx 0.7$. The frequency of the
SED peaks of both the synchrotron and EC emission should obey
$\nu_{\rm peak}\propto \delta \propto F_\gamma^{1/3}$. 
\item{} An increase in bulk Lorentz factor $\Gamma$, and therefore Doppler factor,
raises both the beaming and $u'_{\rm seed}$, as well as $\nu_{\rm peak}$ of both the
synchrotron and EC SEDs. The increase in the seed photon field actually decreases the
synchrotron luminosity in the plasma frame, since EC scattering then consumes a higher 
fraction of the electron energies. If the plasma-frame EC luminosity is already dominant, 
then it will not increase much, since the electrons were already expending nearly all of
their energy on EC emission. We then derive the rough dependence
$F_{\rm opt}\propto \delta^{4.6}{u'_{\rm seed}}^{-1} \propto \delta^{4.6}\Gamma^{-2} \propto \delta^{2.6}$
if we approximate that $\delta\propto \Gamma$. Since $F_\gamma\propto \delta^3$, we
obtain $\zeta\approx 1.2$. 
\item{} An increase solely in the magnetic field strength $B$ by a factor $f_B$ would cause
the synchrotron flux to rise by a factor $f_B^{1+\alpha_{\rm opt}}$ while the frequency of
the peak of the synchrotron SED increases by a factor of $f_B$. The EC flux would 
decrease slightly owing to the higher fraction of electron energy that goes into
synchrotron radiation. In this case, $\zeta$ would be a small negative number.
\item{} An increase only in the energy density of seed photons $u_{\rm seed}$ by a factor
$f_{\rm seed}$ -- either because of changes within the source of the photons or because 
of a shift in position of the radiating plasma toward the source of seed photons -- 
would cause the ratio of EC to synchrotron flux to rise in 
proportion to $f_{\rm seed}$, while the frequencies of the synchrotron and SSC peaks
would remain constant. Without a change in the number of radiating electrons, the
$\gamma$-ray flux would increase only slightly, but the synchrotron flux would decrease
by a factor of $f_{\rm seed}^{-1}$. This would result in $\zeta\ll -1$, with the exact
value changing with the level of dominance of the EC luminosity compared with the
synchrotron power. If the number of electrons also increases to create a flare, then
a high positive value of $\zeta$ is possible.
\end{enumerate}

The value of $\zeta\approx 1$ during time interval (1) of Figure \ref{opt_gamma} agrees
with scenario (i), but the higher frequency of the peak in the $\gamma$-ray SED
(see Fig.\ \ref{3c279sed}) at higher flux levels suggests that cases (ii) or (iii) might
apply despite their respectively lower (0.7) and higher (1.2) values of $\zeta$. The
very high value of $\zeta$ and pronounced variations during interval (2) agree with the 
expectation of a combination of scenarios (i) and (v). This suggests that the sites of the
high-amplitude flares during this period also involved steep gradients in the energy
density of seed photons. In concert with our finding that emission-line fluxes in
3C~279 are proportional to optical (and presumably UV) continuum fluxes, with
a time delay $<2$ months, this might be possible if the seed photons are stimulated
by the synchrotron flares, providing more seed photons for inverse Compton scattering
during subsequent flares.

During time interval (3) of Figure \ref{opt_gamma}, which is limited to the 
highest-amplitude optical outburst, the value of $\zeta\approx 0.8$ is closest to
that expected for scenario (iii), but this is contradicted by the relatively small
$\gamma$-ray to optical flux ratio, $\lesssim4$. The latter suggests either a rise
in the magnetic field strength or a decrease in $u'_{\rm seed}$. These considerations
imply that an increase in both the magnetic field and number of radiating electrons,
probably combined with a decrease in the seed photon field, occurred during interval (3),
such that $\zeta$ adopted an intermediate value between cases (i), (iv), and (v).

The value $\zeta \approx 2$ during time interval (4) corresponds to a time of very high
EC dominance (see Fig.\ \ref{3c279sed}). Perhaps a combination of scenarios (iii) and (iv) --
changes in both bulk Lorentz factor and seed photon energy density -- could give
the observed value of $\zeta$.

\subsubsection{Radio-band correlations}\label{radio_corr}

Visual inspection of Figures \ref{lc_UV_NIR} and \ref{lc_radio} reveals that there are few common details in the optical and radio light curves that would allow one to test the possible existence of correlations and time lags between the two. If we compare the 
optical $R$-band and radio 250~GHz light curves, both of which have good sampling, we find only two intervals of common elevated flux level, TJD~54000--54300 and TJD~55800--56100, and a dip over TJD$\sim$55000--55300. However, close correspondence between different light curves as one considers lower radio frequencies, with time lags between them, are clearly visible. Before evaluating the time lags, we consider what the results of our
analysis would be if we did not possess data at frequencies from 22 to 8~GHz. In this case, we would associate maxima observed at 350--37~GHz close to TJD~55100 with the maximum observed at 5~GHz on TJD$\sim$56800, deriving a delay between variations in these bands of $\sim700$ days. This is certainly a mis-identification, as we readily see when including the 22--8~GHz data. The dashed slanted lines in Fig.~\ref{lc_radio} give a hint of the `real' values of the delays, which are (across the widest span of frequencies) no more 
than 300 days.

The progressive smoothing with decreasing frequency and, eventually, the disappearance of the most prominent maximum at TJD~56100, are naturally explained by the increased volume of the jet at lower frequencies, which smooths out the fine structure of variability. However, this does not explain the appearance and progressive growth of the emission feature first seen at 100~GHz close to TJD~56600 and culminating at 5~GHz near TJD~56800. A possible explanation of these two, markedly different, kinds of behaviour, is the same as suggested in \citet{Raiteri2017}: that of a twisted inhomogeneous jet. Within this approach, magnetohydrodynamic instabilities or rotation of the twisted jet cause different jet regions to change their orientation and hence their relative Doppler factors.

\subsection{Jet kinematics}\label{vlba}
\begin{figure*}
\begin{center}
   \includegraphics[width=\linewidth,clip]{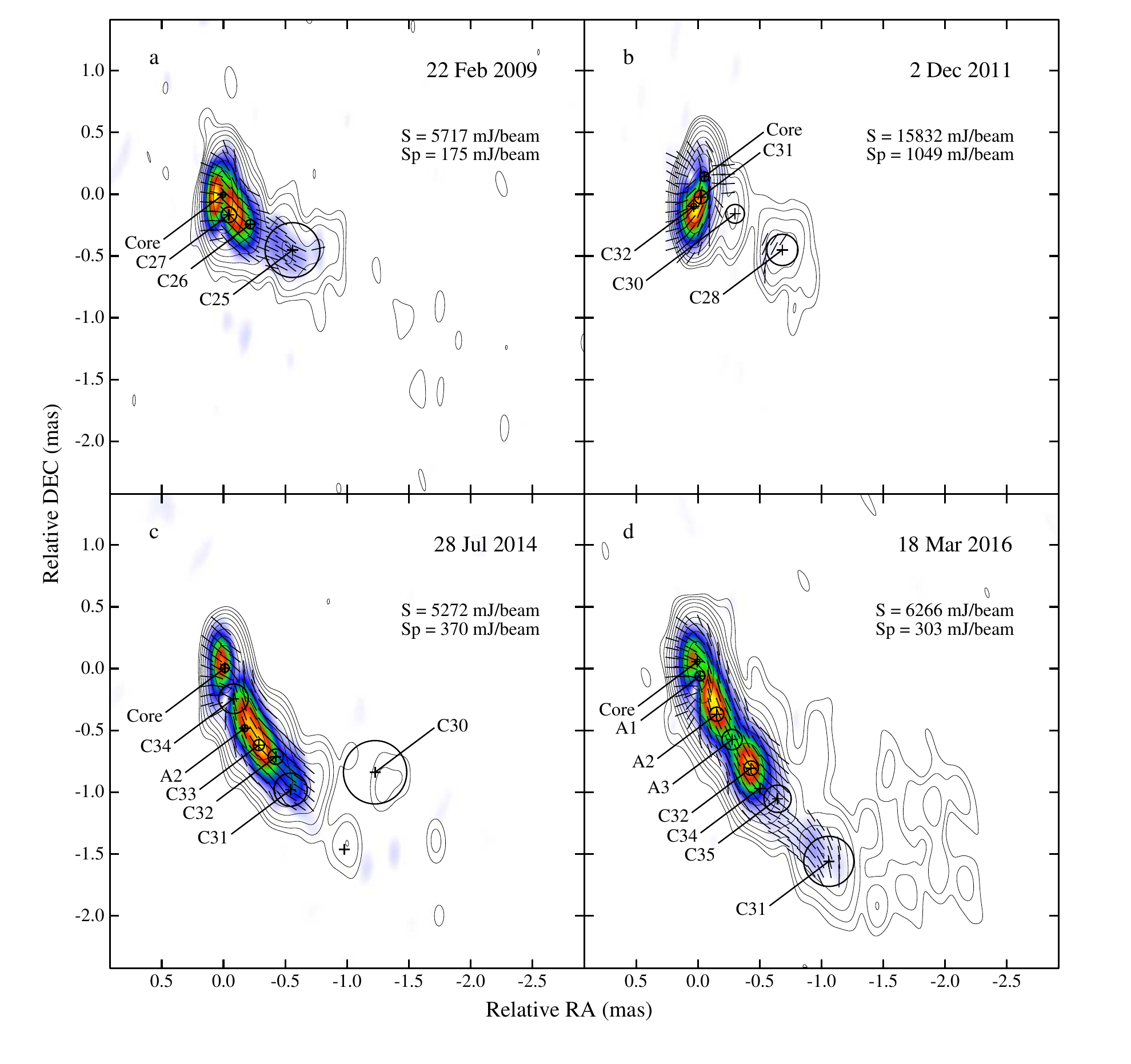}
	\end{center}
   \caption{A set of VLBA images at 43~GHz in total (contours) and polarized (colour scale) intensity ($S$ and $Sp$, correspondingly) overlaid with fitted Gaussian circular components; the images are convolved with a beam of 0.36$\times$0.15~mas$^2$ at PA=-10$\degr$; linear segments within colour scale show direction of EVPA; four different epochs are chosen to show significant changes in the jet structure as discussed in \S\,\protect\ref{vlba}. 
 }
\label{vlbaimages}
\end{figure*}

\begin{figure}
\begin{center}
   \includegraphics[width=\columnwidth,clip]{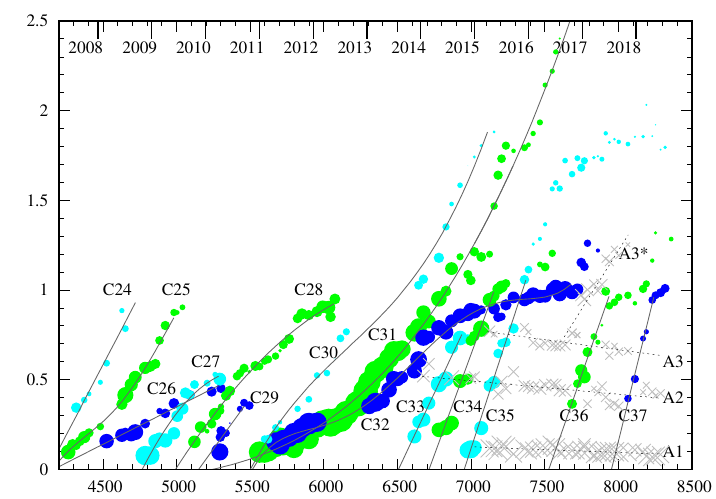}
        \caption{Separation of jet features from the core, A0, during 2007-2018, partly adopted from \protect\citet{Jorstad2017}. 
        Each line or curve represents a polynomial fit to the motion of the respective knot. The diameter of each symbol 
        is proportional to the logarithm of the flux density of the knot, as determined by model fitting of the VLBA data.}
\label{movements}
\end{center} 
\end{figure}

Figure~\ref{vlbaimages} shows  the total and polarized intensity VLBA images of 3C~279, convolved with an elliptical Gaussian beam that approximates the angular resolution
at epochs when all 10 VLBA antennas operated. We follow the historical designation 
of moving components at 43 GHz started by \citet{Unwin1989, Wehrle2001}, and continued  by \citet{Jorstad2004, Jorstad2005, 
Chatterjee2008, Larionov2008}, and \citet{Jorstad2017}. During 2007-2018 we identify 14 moving knots in the jet, C24-C37 (see Fig. \ref{movements}). Knots C24-C32 correspond to 
those identified by \cite{Jorstad2017}.
The apparent speed of the moving components, $\beta_{\rm app}$, has been determined using the same procedure as 
defined in \cite{Jorstad2005}. Since almost all motions in 3C~279 are non-ballistic, in order to derive the most accurate 
values of the time of passage of a knot through the VLBI core, $T_0$\footnote{$T_0$ is the time when the centroids of the knot and 
the core coincide, also called the `ejection' time.}, we use only those epochs when a component is within 1 mas of the core, 
inside of which its motion is assumed to be ballistic. Knots C30, C31, C32, C33, C34, and C35 can be associated with features 
C1, C2, C3, NC1, NC2, and NC3, respectively, in \cite{Rani2018}. Table~\ref{kinematics} shows angular, $\mu$ (in mas/yr), 
and apparent, $\beta_{\rm app}$ (in units of $c$), velocities, the ejection time, $T_0$,
the average projected position angle with respect to the core, $\langle \Theta \rangle$, the average distance from 
the core when the knot is observed, $\langle R \rangle$, and the average flux density, $\langle F \rangle$, of knots ejected between 2007 and 2018 
(partly adopted from \cite{Jorstad2017}). The derived apparent velocities are in the range from 5 to 37$c$. In addition 
to the core A0, we find three quasi-stationary components, A1, A2, and A3, located at $\sim$ 0.1, 0.5, 0.7 mas from the core,
respectively. These knots appear in the jet after the ejections of very bright
knots C31 and C32. Knots C30 - C37 exhibit especially rapid motions, with apparent 
velocities ranging from 20 to 37$c$. Components C36 
and C37 decelerate significantly after reaching a distance of about 1~mas from the core. Knots C30 and C31 possess velocities up to $\sim20c$ at distances $>$1.5~mas, factors of 
$\sim$ 2 and 3 times higher than those near the core, 
respectively. Table~\ref{crossing} lists times at which moving knots C35, C36, and C37 cross stationary components A1, A2 and 
A3, according their kinematics.

\begin{table*}
\caption{\bf Jet structure and kinematics}
\label{kinematics}
\begin{tabular}{c | c | c | c | c | c | c | c | c |}
\hline

Knot & N & $\mu$(mas/yr) & $\beta_\mathrm{app} $ & $T_0$(TJD)  & $T_0$(year)  & $\langle\Theta\rangle(\degr)$ & $\langle R \rangle$(mas) & $\langle F \rangle$(Jy)\\  
\hline

C24    &  7  & $0.503\pm0.031 $ & $16.01\pm0.99 $ & $54064\pm 40 $ & $2006.90\pm0.11 $ & $-125.4\pm2.8$ & $0.560\pm0.210 $ & $0.43\pm0.22$\\
\hline
C25    &  17  & $0.386\pm0.014 $ & $12.30\pm0.43 $ & $54133\pm 33 $ & $2007.09\pm0.09 $ & $-124.9\pm13.1$ & $0.410\pm0.250 $ & $1.13\pm0.65$\\
\hline
C26    &  13  & $0.153\pm0.015 $ & $4.87\pm0.48 $ & $54137\pm 51 $ & $2007.10\pm0.14 $ & $-139.6\pm6.1$ & $0.290\pm0.100 $ & $1.09\pm0.77$\\
\hline
C27    &  14  & $0.358\pm0.021 $ & $11.39\pm0.66 $ & $54754\pm 29 $ & $2008.79\pm0.08 $ & $-150.6\pm12.1$ & $ 0.310\pm0.170$ & $ 1.77\pm1.65$\\
\hline
C28    &  30 & $0.309\pm0.008 $ & $9.83\pm0.26 $ & $54933\pm 47 $ & $2009.28\pm0.13 $ & $-131.5\pm6.8$ & $0.600\pm0.240 $ & $0.81\pm0.37$\\
\hline
C29    &  8  & $0.413\pm0.063 $ & $13.15\pm2.00 $ & $55149\pm 55 $ & $2009.87\pm0.15 $ & $-122.4\pm5.1 $ & $0.250\pm0.100$ & $0.91\pm1.17 $\\
\hline
C30    &  21(12)  & $0.394\pm0.018 $ & $12.56\pm0.57 $ & $55452\pm 22 $ & $2010.70\pm0.06 $ & $-139.3\pm10.3$ &$0.400\pm0.230 $ & $0.42\pm0.14$\\
\hline
C31    &  55(23)  & $0.227\pm0.012 $ & $7.23\pm0.40 $ & $55539\pm 77 $ & $2010.94\pm0.21 $ & $-181.2\pm19.3$ & $0.320\pm0.200 $ & $11.29\pm5.61$\\
\hline
C32    &  48(10)  & $0.168\pm0.016 $ & $5.35\pm0.53 $ & $55368\pm 84 $ & $2010.47\pm0.23 $ & $-175.9\pm18.3$ & $0.320\pm0.140 $ & $4.61\pm2.24$\\
\hline
C33     &9(8)   & $0.655\pm0.009  $ & $20.86\pm0.29 $ & $56507\pm29 $ & $2013.59\pm0.08 $ & $-163.8\pm6.5$ & $ 0.446\pm0.186$ &$2.01\pm0.49$ \\
\hline
C34     &16(10)  & $0.83\pm0.03  $ & $26.1\pm0.9 $ & $56711\pm6 $ & $2014.15\pm0.02 $ & $-157.3\pm6.1$ & $0.77\pm0.33$ & $1.39\pm0.78  $\\
\hline
C35     &38(19)  & $0.924\pm0.006  $ & $29.43\pm0.18 $ & $56953\pm22 $ & $2014.81\pm0.06 $ & $-151.7\pm6.5 $ & $1.347\pm0.604$ & $0.68\pm0.93$\\
\hline
C36     &19(9)   & $0.807\pm0.018  $ & $25.71\pm0.57 $ & $57511\pm11 $ & $2016.34\pm0.03 $ & $-150.5\pm2.8 $ & $0.859\pm0.266 $ & $0.56\pm0.34 $\\
\hline
C37     & 8(6)   & $1.161\pm0.034  $ & $36.97\pm1.08 $ & $ 57937\pm7$ & $2017.51\pm0.02 $ & $-148.3\pm2.0$ & $0.788\pm0.234$ & $0.51\pm0.13$\\
\hline
A1     & 34     & -  &  - &  - & - & $-164\pm7$ & $0.10\pm0.02$ &$4.09\pm1.38$ \\ 
\hline
A2     & 13     & -  &  - &  - & - & $-158\pm2$ & $0.46\pm0.03$ &$1.17\pm0.16$ \\ 
\hline
A3     & 18     & -  &  - &  - & - & $-156\pm3$ & $0.71\pm0.04$ &$0.63\pm0.23$ \\
\hline

\end{tabular}
\end{table*}

\begin{figure}
\begin{center}
   \includegraphics[width=\columnwidth,clip]{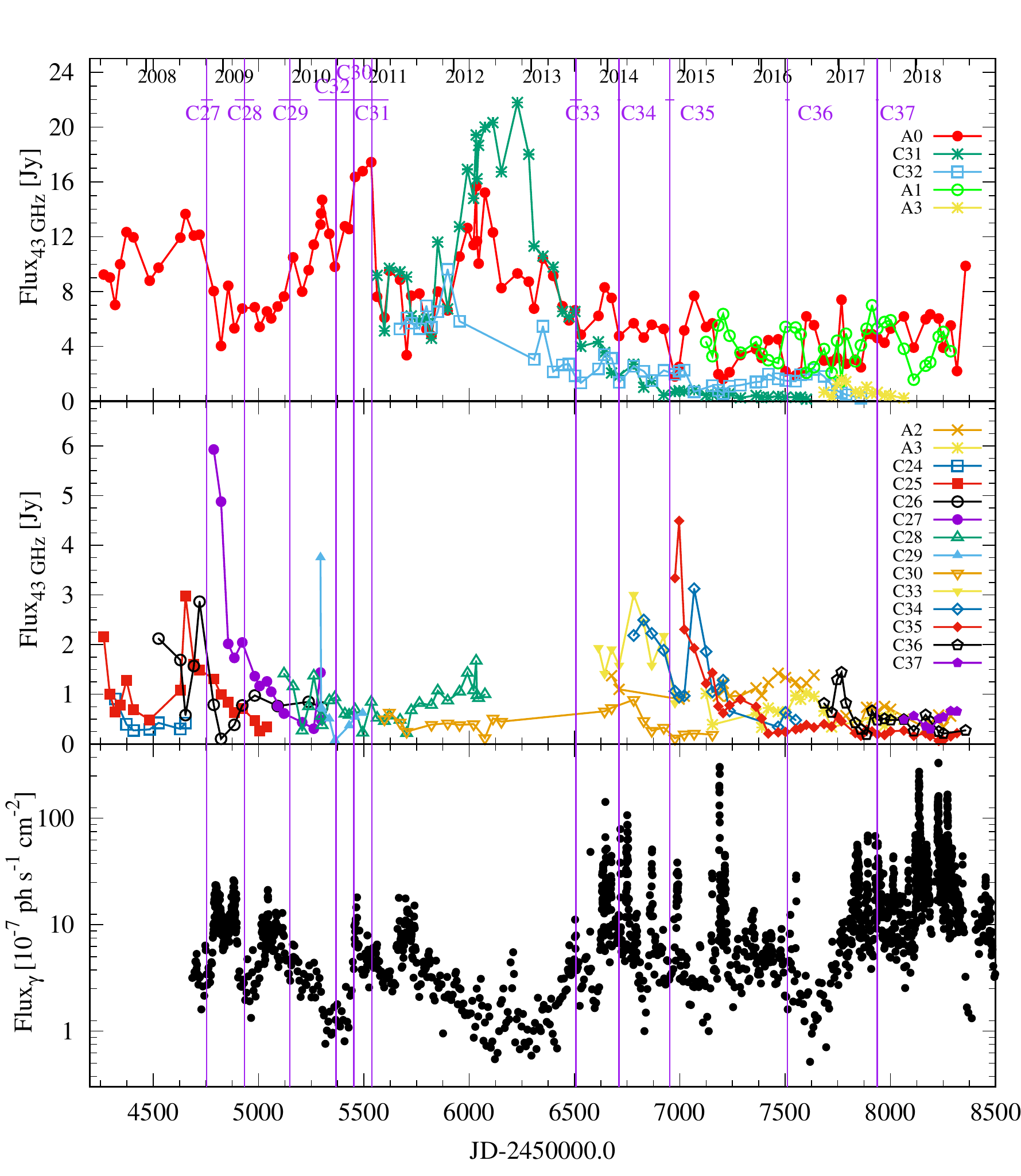}
        \caption{{\it Top}: 43 GHz light curves of the core, A0, stationary feature, A1, and brightest moving knots in the jet, C31 \& C32.
         {\it Middle}: 43 GHz light curves of other knots detected in the jet during 2007-2018. {\it Bottom}: $\gamma$-ray light curve. The vertical lines show the times of 
         ejection of moving knots.}
\label{lc_knots}
\end{center} 
\end{figure}

The ejection of each component appears to be associated with a flare in the core (see Figure \ref{lc_knots}). Components 
ejected after extremely bright knot C31, with initial projected position angle $\Theta\sim$-210$\degr$ \citep{Jorstad2017}, 
follow a new jet direction of $\sim-160\degr$, in contrast to the usual direction of the parsec-scale jet 
of $\sim−130\degr $\citep{Jorstad2005,Lister2016}. Figure~\ref{optical_radio_polar} (bottom panel) shows the behaviour of 
the position angle (PA) of the inner jet (within 0.3~mas from A0). The PA undergoes significant changes during 2008-2018 
(from $\sim-100\degr$ to $-200\degr$), with especially dramatic variations between 2009 and 2012. According to Table~\ref{kinematics},
this period is associated with ejection of 5 knots, including the brightest features, C31 and C32,
and extreme variability of the flux density of A0 (Figure \ref{lc_knots}).
After 2013 the core has lower average flux density than before at 43 GHz, which might be 
related to changes of the inner jet direction.

\begin{table} 
\caption{\bf Stationary components crossing times and nearby $\gamma$-ray flares.}
\label{crossing}
\begin{tabular}{c | c | c | c | c | c|}
\hline
(1) & (2) & (3) & (4) & (5) & (6)\\
\hline

C35& A1 &$0.10\pm0.02$ & $56992\pm8 $& 56998 & 38\\ 
\hline
& A2 &$0.46\pm0.03$  & $57134\pm13 $ & 51153 & 30\\ 
\hline
& A3 &$0.71\pm0.04$ & $57233\pm18 $ & 57214  & 45\\
\hline
C36& A1 &$0.10\pm0.02$ & $57556\pm11$ & 57552 & 30\\ 
\hline
& A2 &$0.46\pm0.03$  & $57719\pm18$ & -- & --\\ 
\hline
& A3 &$0.71\pm0.04$  & $57832\pm26$ & 57842 & 56\\
\hline
C37& A1 &$0.10\pm0.02$ & $57968\pm7 $  & 57928 & 68\\ 
\hline
& A2 &$0.46\pm0.03$  & $58082\pm15$ & 58118 & 69\\ 
\hline
& A3 &$0.71\pm0.04$  & $58162\pm19$ & 58136 & 249\\
\hline
\end{tabular}\\
(1) -- Moving knot\\
(2) -- Stationary component\\
(3) -- Size of stationary component, $\langle R \rangle$(mas)\\
(4) --  $T_\mathrm{cross}$(TJD)\\
(5) --  $T_{\gamma}$(TJD) within $\pm 40$ days from $T_\mathrm{cross}$ \\
and $F_\gamma > 20 \cdot 10^{-7}$ ph cm$^{-2}$ s$^{-1} $ )\\
(6) -- $\gamma$-ray flux density ($10^{-7}$ ph cm$^{-2}$ s$^{-1} $ )\\
\end{table}

Based on the properties of the components, we can distinguish four different periods of 
jet activity that can be associated with changes in the jet structure: \\
\begin{itemize}
\item[(a)] 2008--mid-2010: four knots (C27-C30) with comparable apparent speeds of
12--13$c$ are ejected,\\$\langle \beta_{app} \rangle=11.7\pm1.5$, 
$\langle \Theta \rangle=-136^{\circ}\pm12$,  $\langle F \rangle=0.98\pm0.57$ Jy;
\item[(b)] late-2010-2013: two extremely bright knots (C31 and C32) with slower apparent 
velocities are ejected, \\ $\langle \beta_{app} \rangle=6.29\pm0.94$,  $\langle \Theta \rangle=-179^{\circ}\pm3$,  $\langle F \rangle=7.95\pm3.34$ Jy, with trajectories
that bend from south-southeast to south-southwest;
\item[(c)] 2013.6--2015: four knots (C33-C36) with similarly high apparent speeds, $\sim20$--$30c$, are ejected, \\ $\langle \beta_{app} \rangle=25.5\pm3.5$,  $\langle \Theta \rangle=-154^{\circ}\pm4$,  $\langle F \rangle=1.02\pm0.47$ Jy ); and 
\item[(d)] late 2015--2018: three stationary knots (A1-A3) appear, and fast knot C37 is 
ejected in mid-2017, \\ $\beta_{app}=37.0\pm1.1$, $\langle \Theta \rangle=-148\degr\pm2$,   $\langle F \rangle=0.51\pm0.13$ Jy). 
\end{itemize}
Figure\,\ref{vlbaimages} illustrates the aforementioned structural changes in the jet. Four epochs were selected to highlight corresponding periods (a), (b), (c), and (d).
The trajectories of the knots reveal changes in the PA of the inner jet (see also Figure~\ref{optical_radio_polar}), or at least the PA of the portion of the jet
cross-section that is involved in the main emission. Particularly striking is the shift in
direction that occurs in late 2010, based on the south-southeastern motion of knots C31 
and C32 (see also \citet{Lu2013} and \citet{Jorstad2017}) and then a swing back toward 
$-154^\circ$, $-18^\circ$ from the previous position angle.

The observed degree of polarization at 43 GHz is quite high during 2007-2018, above 10\% nearly the entire time. 
The optical EVPA is in good agreement with the 43 GHz EVPA, and both are roughly parallel to the direction of the jet 
(see Figure~\ref{optical_radio_polar}). Figure~\ref{optical_radio_polar} also shows a remarkable swing of the EVPA at radio 
wavelengths and optical wavelengths, which occurs simultaneously with the dramatic change in the inner jet direction mentioned above.  

\subsection{Relationship between multiwavelength variability and changes in jet structure}
\label{interpretation}

Although the variations that we observe in multiwavelength
flux, polarisation, and structure of the jet are quite complex, we offer the following
approximate interpretation of changes in the physical processes that dominate at different 
epochs:

\begin{enumerate}
\item[(1)] TJD~54700-55400 (late 2008 - mid-2010): As discussed in \S\ref{gamma-optical}, 
the frequency of the EC SED peak (and probably of the synchrotron peak) is higher when 
the flux is greater (see Fig.\ \ref{3c279sed}). This trend, plus the near-unity slope of 
the $F_\gamma$ vs.\ $F_R$ relation (Fig.\ \ref{opt_gamma}), can be explained if the 
variations are dominated by changes in the Doppler factor because of a shift in viewing 
angle (see scenario (2) of \S\ref{sect:sed}). The factor of $\sim 10$ and 20 
decline in the $\gamma$-ray and $R$-band fluxes, respectively, over the time period is 
consistent with a Doppler factor decrease by a factor of 1.7. This qualitatively
conforms with the relatively moderate apparent speeds of 10-13$c$ of radio knots during this time span (see Table \ref{kinematics}) if the viewing angle
was greater than the value of $\sim\Gamma^{-1}$ that maximizes superluminal motion.
In fact, detailed analysis of knots C27-29 by \citet{Jorstad2017} determines that
the viewing angle indeed increased from $\lesssim 2\degr$ to $\sim 6\degr$ from late
2008 to late 2009 as the optical and $\gamma$-ray fluxes dropped to very low values.
\item [(2)] TJD~55500-57600 (late 2010 - mid-2016): As discussed  in 
\S\ref{gamma-optical}, the high value of the slope of the $F_\gamma$ vs.\ $F_R$ 
relation (Fig.\ \ref{opt_gamma}) plus the high-amplitude variations imply that the seed 
photon density changes across the region where the electrons are accelerated [a 
combination of scenarios (1) and (5) of \S\ref{sect:sed}]. (We note that during
the first section from late 2010 to mid-2011, the optical and
$\gamma$-ray fluxes rise by about a similar factor of $\sim5$, which is consistent with
an increase solely in the number of radiating electrons.) This is the same time
period when two very bright knots are ejected in a very different direction than
at earlier and later epochs. From late 2013 
until late 2015, the X-ray and $\gamma$-ray flux variations have very high amplitudes
compared with earlier epochs, with $\gamma$-ray doubling times as short as 5 min
\citep{Hayashida2015,Ackermann2016}. The beginning of this behaviour coincides with the
resumption of ejections of new knots after a $\sim2$-yr hiatus as the optical, X-ray,
and $\gamma$-ray activity becomes strong again. These knots have quite high apparent 
speeds, in the 
$20$-$30c$ range. The establishment of stationary features A1, A2, and A3 in late 2015
might create structures (e.g., standing shocks) that both accelerate electrons and provide sources of (synchrotron) seed photons for IC scattering. 
\item [(3)] TJD~57700-57850 (late 2016 - early 2017): During this relatively brief period, 
the slope of the
$F_\gamma$ vs.\ $F_R$ relation (Fig.\ \ref{opt_gamma}) reverts to a value near unity.
The synchrotron flux rises by a factor $\sim20$ to its highest level of the 10 years of
our study, while 
the Compton to synchrotron flux ratio is modest, $\sim4$. Knot C37, which crosses the 
millimetre-wave core shortly (0.25 yr) after this time interval ended, moves at the 
highest apparent speed observed in our study, $37c$. We can explain the behaviour of the
optical and $\gamma$-ray flux qualitatively during this period by an increase in both 
the magnetic field strength and number of radiating electrons, perhaps accompanied by
a decrease in the energy density of seed photons.
\item [(4)] TJD~57850-58350 (early 2017 - mid-2018): During this time interval, the slope 
of the $F_\gamma$ vs.\ $F_R$ 
relation (Fig.\ \ref{opt_gamma}) switches to $\zeta\sim2$. The extremely high
apparent speed of knot C37 indicates that the Lorentz factor and/or direction of motion
indeed changes during this time span. Variations in the
external seed photon energy density by the optical outburst in 2017 -- scenario (4) in 
\S\ref{sect:sed} -- might steepen the slope $\zeta$ from 1.2 to the observed value of 
1.9.
\end{enumerate}

If the stationary features A1-3 supply seed photons for EC emission by electrons in
the superluminal knots, flares of high-energy photons should occur contemporaneously with the crossing of knots through the features. Comparison of the times of the crossings
listed in Table \ref{crossing} with times of $\gamma$-ray flares (none of which were missed, since there are no gaps in coverage) finds a $\gamma$-ray flare within 40
days of the epoch of knot crossing in: all three passages of knots through A1; two out
of three through A2; and all three through A3, as seen in Figure~\ref{lc_knots}. However, while this suggests
consistency with the hypothesis that stationary emission features in the jet supply
seed photons for many of the flares, all but two of the above crossing-flare pairs
occur during a period when there were so many closely-spaced flares that a coincidence
is guaranteed.

\citet{Ackermann2016} find that a standard EC model matching the observed characteristics
of the $\gamma$-ray flux doubling time scale of 5 min during a flare in June 2015 requires
a Lorentz factor $\Gamma>50$, with $\Gamma\sim 120$ needed to raise the derived energy density of the magnetic field to equipartition with that in relativistic electrons. Our measurement of an apparent speed of $37c$ in 2017 requires $\Gamma>37$, and the value required to explain the proper motion of knot C35 in 2015 is only 20\% lower. If turbulent 
motions are superposed on this systemic velocity, then local values of $\Gamma\sim70$ 
would be possible if the maximum turbulent velocity were the relativistic sound speed of
$c/\sqrt{3}$, and even higher for supersonic turbulence or `mini-jets' formed via
magnetic reconnections \citep{Giannios2009}.

\subsection{Polarimetric behaviour at optical and radio wavelengths}\label{pol_behaviour}

\begin{figure}
   \centering
  \includegraphics[width=\columnwidth,clip]{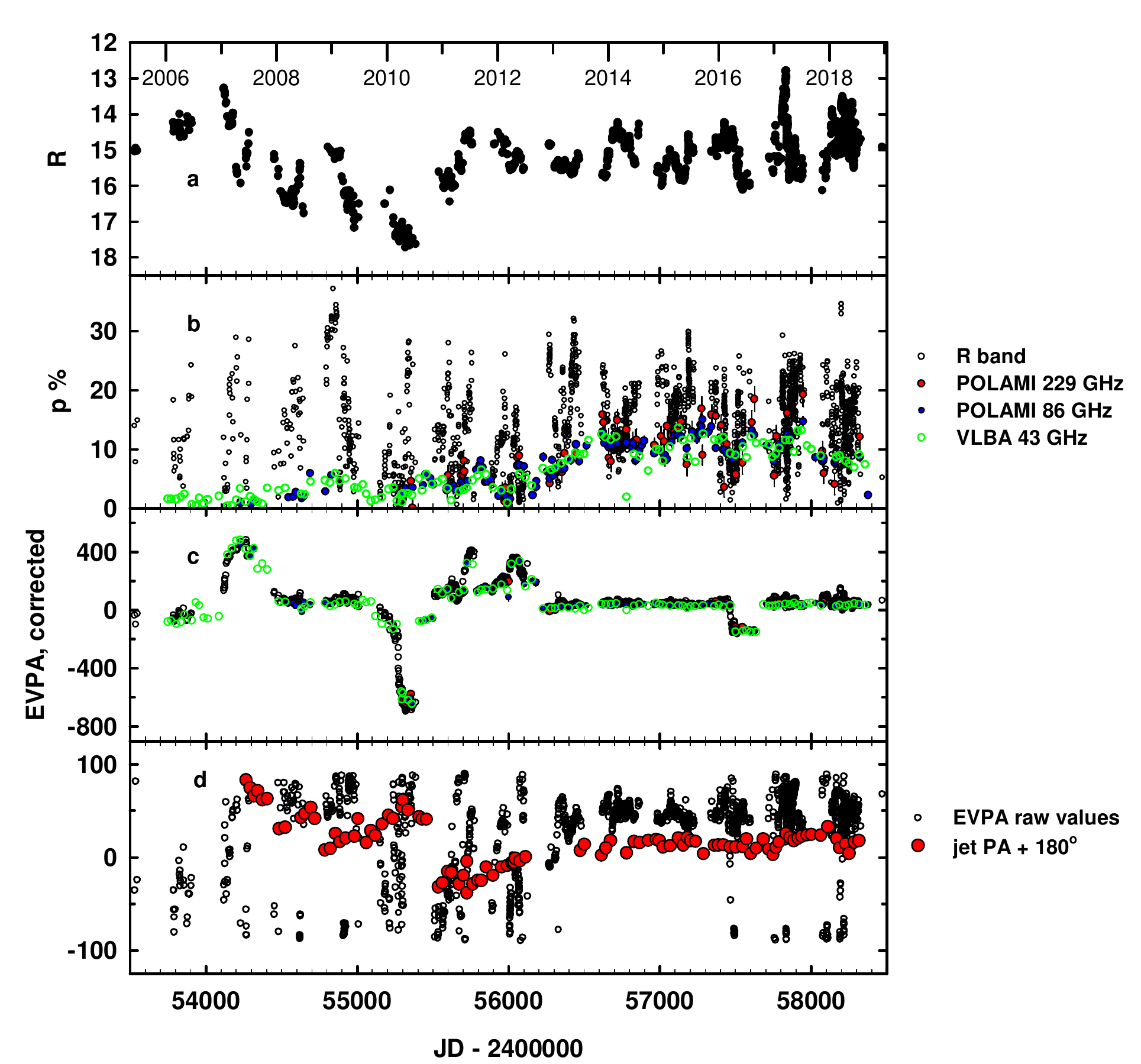}
      \caption{
			(a) $R$-band light curve of 3C~279; (b) fractional polarization at optical and radio bands; (c) electric vector position angle (EVPA) at optical and radio bands, with 180\degr ambiguity solved; (d) raw values of optical EVPA (black) and position angle of the 43 GHz jet within 0.3~mas of the core (red). 
			}
         \label{optical_radio_polar}
   \end{figure}

Figure \ref{optical_radio_polar} shows the time dependencies of optical $R$ band magnitude, degree of polarization, and electric vector position angle (EVPA) of 3C~279. The evolution of polarization parameters is shown for optical data and for three radio bands: 230, 100 and 43~GHz. We solve the $n\times180\degr$ ambiguity between optical and less well-sampled radio data that arise by minimizing the difference between the radio and optical EVPA values. We note the occurrence of exceptionally high-amplitude changes of the polarization degree (PD), from nearly 0\% to $>30\%$. 

This high level of scatter of the optical PD is in striking contrast to the relatively smooth behaviour of the PD at radio frequencies; in fact, the radio PD mostly represents the lower envelope of the optical PD, as is seen in Figure \ref{optical_radio_polar}(b). We propose that this contrast between optical and radio PD scatter is a consequence of very different volumes radiating at the two wavelength ranges in a plasma with a strong
turbulent component of magnetic field: at radio frequencies the number of radiating cells with different magnetic field orientations is substantially larger than at optical bands, and the effect of each cell is diminished after vector co-addition of polarized intensity from all of the cells in the radiating volume. However, as seen in Figure \ref{optical_radio_polar}(c), there is almost no difference in the direction of the electric vector position angle (EVPA) between optical and radio frequencies. At most epochs, this direction is parallel to the jet axis, strongly implying that a common process partially orders the magnetic field across the entire wavelength range, such as 
shocks or a helical field component. This ordering, superposed with the more chaotic field 
structure indicated by variability of the PD and sometimes the EVPA, agrees well with
scenarios like the turbulent, extreme multi-zone model (TEMZ) developed by \citet{Marscher2014}.

In Figure \ref{optical_radio_polar}\,(d) we see a persistent offset between the mean optical
EVPA and inner ($\lesssim0.3$~mas) jet direction of $\sim30^\circ$ starting in 2013.
We note that the EVPA during this time interval is similar to the downstream (0.5-1.5~mas)
jet direction of knots ejected prior to 2010 (see Table \ref{kinematics}). This implies
that the inner jet direction after mid-2013 might correspond to an elongated emission
feature (e.g., a site of magnetic reconnections) oriented at an angle to the general jet
flow. Given the strong projection effects, this angle can be as small as the angle at
which we view the jet axis, $1.4$-$6^\circ$.

\begin{figure}
   \centering
  \includegraphics[width=\columnwidth,clip]{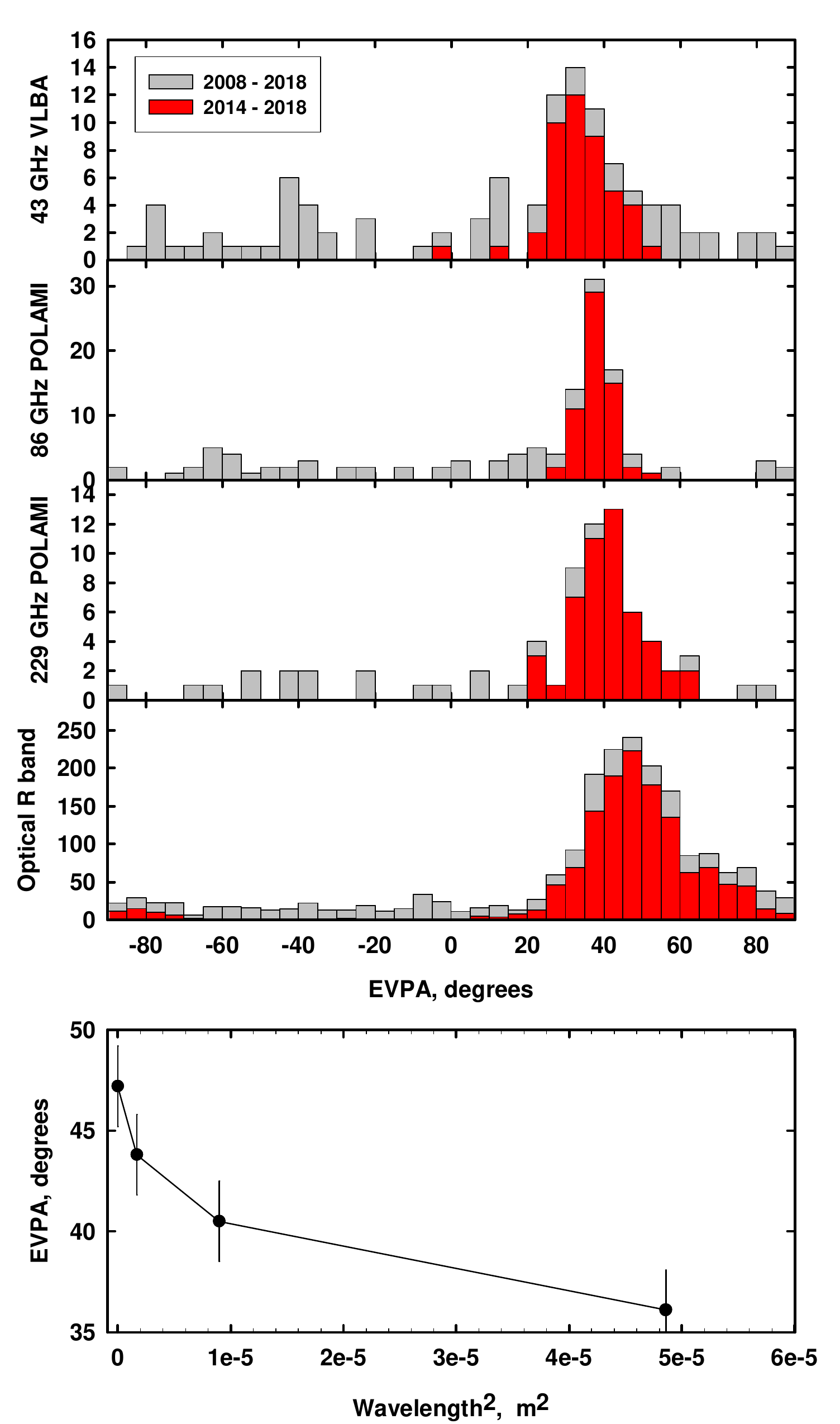}
      \caption{
			Top: Histograms of the EVPA distribution in optical and radio bands. Bottom: Positions of maxima of the 2014--2018 distributions vs. square of wavelength. 
			}
         \label{histogram}
   \end{figure}

The large number of polarimetric measurements both in optical and radio bands allow us to look for minor differences between the EVPA behaviour at different wavelengths. To do this, in Fig.~\ref{histogram} we plot histograms of the EVPA distributions in optical $R$ band and radio frequencies of 229~GHz, 86~GHz, and 43~GHz. Unlike the plot in Fig.~\ref{optical_radio_polar}\,(c), we constrain the range of EVPAs to $[-90, 90]$ degrees. We immediately see a monotonic shift of the positions of maxima in the histograms with wavelength. We interpret this shift as a sign of Faraday rotation, and plot the values of these maxima relative to the square of wavelength in the bottom panel of the same figure. The values of the rotation measure (RM) obtained in this way for the pairs of frequencies $10^{5.6}$--229~GHz, 229--86~GHz, and 86--43~GHz are, correspondingly, $-35100\pm3000, -7900\pm1000$, and $-1900\pm500$~rad$\cdot$m$^{-2}$. Neither the signs nor frequency dependence of the RM contradict earlier determinations \citep[see, e.g.,][]{Kang2015, Park2018}.

\begin{figure}
   \centering
  \includegraphics[width=\columnwidth,clip]{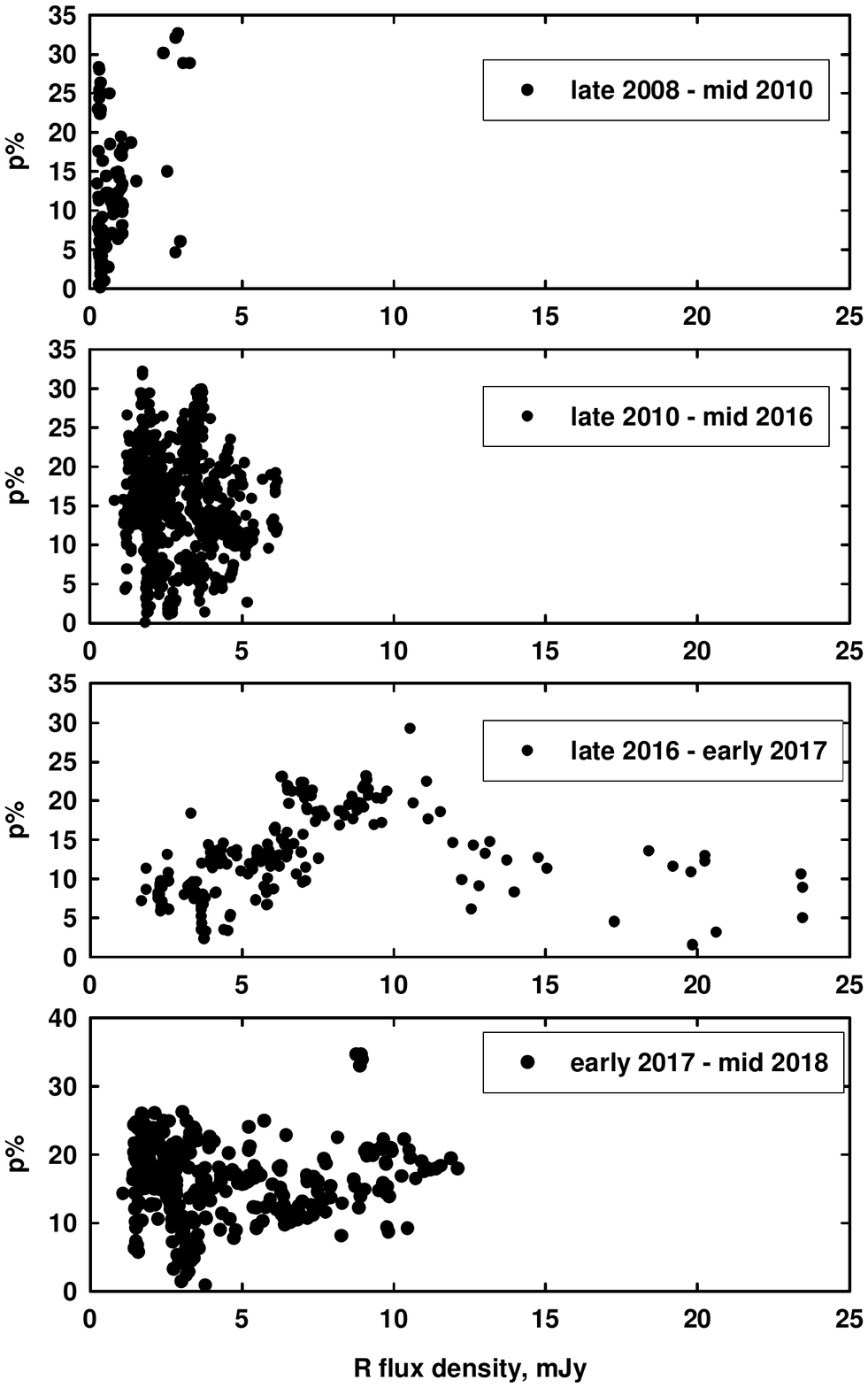}
      \caption{Behaviour of fractional polarization of 3C~279 at different stages of activity.}
         \label{polar_degree}
   \end{figure}

Figure~\ref{polar_degree} shows the dependencies of the optical PD on $R$-band flux density for different stages of activity (1)--(4), as defined  in \$~\ref{interpretation}. The very different types of behaviour of PD vs. flux density may reflect different physical and geometrical conditions in the jet, as discussed above. 
Examples include changes in the Doppler factor due to shifts in viewing angle or variations in magnetic field strength and/or number of radiating electrons. Given the complexity of the observed behaviour, it is not surprising that time-limited campaigns may yield conclusions that are (or seem to be) contradictory.

\begin{figure}
   \centering
  \includegraphics[width=\columnwidth,clip]{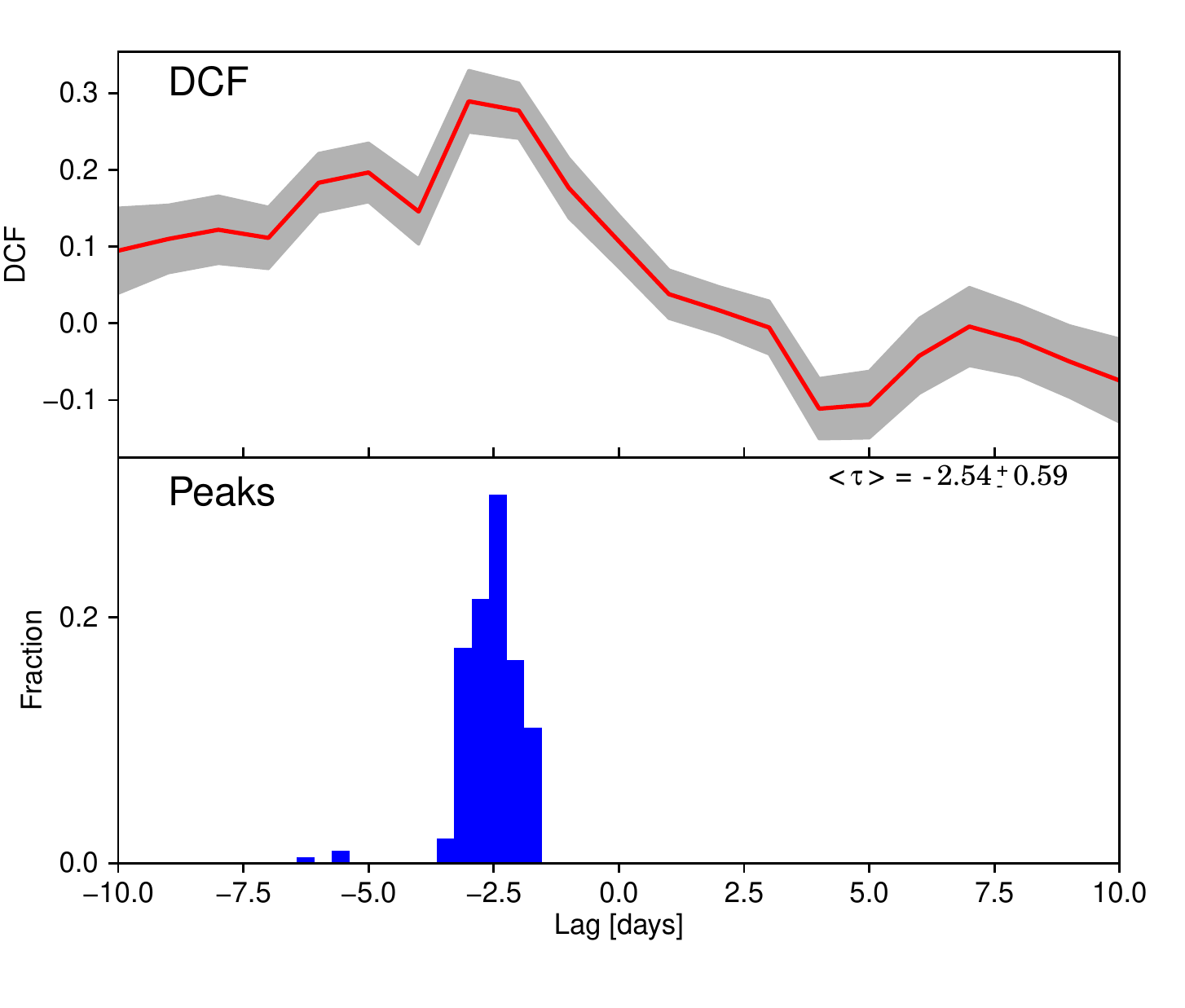}
      \caption{Discrete correlation function between normalized Stokes parameters $q$ and $u$ in optical $R$ band for the time interval 2008--2018 (top) and statistical distribution of the lag of the peak from Monte Carlo simulations (bottom). The grey area in the upper panel correspond to $\pm 1 \sigma$ spread of simulated DCFs.}
         \label{dcfqu}
   \end{figure}

	Further important information that can be obtained from polarimetric data concerns possible spiral structure of the jet or helical structure of the magnetic field. During the past decade, several robust cases have been reported of optical EVPA rotations that mostly occurred close to optical, X-ray, and/or $\gamma$-ray outbursts, e.g., in BL~Lac \citep{Marscher2008}, 3C~279 \citep{Larionov2008}, PKS~1510-089 \citep{Marscher2010}, S5~0716+71 \citep{Larionov2013}, and CTA~102 \citep{Larionov2016a}. However, regular patterns in the behaviour of polarization vector could also occur during quiescent states. \citet{Larionov2016b} suggested using a DCF analysis between normalized Stokes parameters $q$ and $u$ to uncover possible regular rotations that might be hidden by stochastic variability. If monotonic rotation of the EVPA is present during the temporal evolution of the polarization parameters, the curves of $q(t)$ and $u(t)$ must have a systematic shift relative to each other. In the opposite case, when only stochastic variability is present, no systematic shift is expected. Moreover, the direction of any rotation can be obtained from the sign of the DCF slope at lag=0: negative slope corresponds to counter-clockwise rotation and \textit{vice versa}. In Fig.~\ref{dcfqu} we plot the DCF between $q$ and $u$ in optical $R$ band, which shows that during 2008--2018 there is a clear indication of anticlockwise rotation of the EVPA. To estimate the significance of the correlation between Stokes $q$ and $u$ parameters, we have performed a Monte-Carlo simulation
in way similar to the one we used to estimate the significance of the $\gamma$-ray -- optical correlation. The only difference is the
method we used to make a set of synthetic curves. The time dependence of the Stokes parameters does not show well defined
peaks, and cannot be adequately fit by a set of double exponential peaks, as is possible for the $\gamma$-ray and optical curves.
To make synthetic light curves of the Stokes parameters, we approximated them using a damped random walk (DRW) model
\citep{Kelly2009, Zu2013}. We used a JAVELIN code \footnote{\url{https://bitbucket.org/nye17/javelin/}} to obtain parameters of the DRW models
for both $q$ and $u$ data sets and to generate $10^4$ random light curves with the derived parameters. By correlating these synthetic
light curves, we find that the maximum difference of 0.4 between the highest-magnitude 
positive and negative correlations of the observed $q$ and $u$ data sets is reached only in $3\cdot 10^{-4}$ cases. Therefore, we can consider this correlation to be statistically significant at the 0.03 per cent level.

\section{CONCLUSIONS}\label{conclusions}

We have reported the results of decade-long (2008--2018) monitoring of the blazar 3C~279
from $\gamma$-rays to 1~GHz radio frequencies. Our data set
includes \emph{Fermi} and \emph{Swift} data, obtained alongside an intensive
GASP-WEBT collaboration campaign, polarimetric and spectroscopic data
collected during the same time interval, and roughly monthly sequences of VLBA
images at 43 GHz.

By analysing the multi-colour flux dependencies, we have isolated the spectra of the
variable optical to near-IR continuum, which follows a power law with slope ranging
from $-1.47$ to $-1.65$. Our multi-epoch optical spectra reveal changes in \ion{Mg}{ii}
and \ion{Fe}{ii} emission-line flux. The \ion{Mg}{ii} line consists of a component at
the systemic redshift of 3C~279, as well as a second component 3500 km~s$^{-1}$ on the
long-wavelength (red) side. The fluxes of both components are proportional to the optical
continuum flux, although the redder component has a stronger dependence. If the redder
feature is free falling toward the central black hole, we estimate that the
line-emitting clouds lie $\sim0.6$ pc from the black hole. The reverberation of
the changing continuum -- which is synchrotron emission from the jet -- is less than
two months, restricting the location of the clouds to within $\sim10\degr$ of the
jet axis as viewed by the black hole. This polar line emission is likely to be a
significant, variable source of seed photons for `external' Compton scattering
creating X-ray and $\gamma$-ray emission.

The SED contains the two-hump shape typical of blazars. The ratio of the higher peak,
which falls in the $\gamma$-ray range, to the synchrotron peak in the IR, exceeds 100
during some outbursts. An analysis of the peak frequencies of the IR and $\gamma$-ray
humps eliminates synchrotron self-Compton as the dominant $\gamma$-ray emission
process during outbursts unless the magnetic field is extremely low, $\sim0.4$ mG, and
the plasma extremely far from equipartition between the magnetic and particle energy 
densities. Instead, EC emission appears to dominate at most epochs, which is consistent 
with the high $\gamma$-ray to IR-optical flux ratios.

The X-ray and $\gamma$-ray behaviour of 3C~279 are remarkably similar, with
a time lag between the two light curves of $\la 3$ hours, indicating co-spatiality of
the X-ray and $\gamma$-ray emission regions. The relation between the $\gamma$-ray and
optical flux is quite complex, with a variety of slopes during different stages of 
activity. The dependence changes from linear to quadratic to a period when the 
$\gamma$-ray flux varied greatly during much more modest optical variations.
At radio bands we have found progressive shifts of the most prominent light curve features
with decreasing frequency, consistent with the emission peaking as it emerges from the
$\tau=1$ surface. In addition, some fluctuations that emerge in the radio light curve  
disappear with decreasing frequency, which suggests different Doppler boosting of 
different radio-emitting zones in the jet.

From our series of VLBA images at 43 GHz, we have identified 14 bright knots that appear
to move at various superluminal speeds. Some of the trajectories curve, with the
direction of `ejection' close to the `core' changing with time. In 2010 two extremely
bright knots initially emerged to the south-east of the core before veering toward the
usual south-western direction of the $\ga0.5$ mas-scale jet. Periods of different
multi-wavelength behaviour of the light curves correspond to changes in the behaviour
of the compact jet. The formation of three stationary emission features, which
occurred after the direction of ejection became more consistent, and increase in
apparent speed of moving knots up to $37c$ coincide with the highest-amplitude high-energy 
and optical outbursts. The extremely high apparent velocity requires a bulk Lorentz factor
exceeding 37, which is nearly the value of $\sim50$ needed to explain extremely rapid
variations during the highest-amplitude $\gamma$-ray flares in 2013-15. Turbulent
motions at the relativistic sound speed could increase this up to $\sim70$ when a
portion of the plasma moves toward the line of sight relative to the systemic flow.

The degree and position angle of linear polarization at optical and millimetre wavelengths
(single-telescope and VLBA) changes with time, with the most rapid variability occurring
at optical wavelengths. The correlation of variations in the optical $q$ and $u$
Stokes parameters corresponds to that expected in the case of a persistent helical
magnetic field component or spiral motion of the radiating plasma.

The patterns of multi-wavelength variability of the flux, polarization, and structure
of the relativistic jet of 3C~279 change over time. The variations in the ratios of
the peak fluxes of the high-energy and optical-IR portions of the SED can be
ascribed to changes in the ratio of the seed-photon to magnetic energy densities in the plasma frame. Such changes can occur via (1) variations in the bulk Lorentz factor of
the plasma, (2) shifts in location of the site of acceleration of relativistic electrons
relative to the emission-line clouds, especially those lying within $\sim10^\circ$ of
the jet axis, or (3) time-dependent reverberation of emission lines following variations
in the UV continuum from either the jet or accretion disk. General variations of the
flux can correspond to fluctuations in the number of radiating electrons, magnetic field
(for synchrotron and SSC emission only), or Doppler factor, with the last of these caused
by a variation in either the bulk Lorentz factor or direction of the velocity vector
of the emitting plasma. The complexity of the variations in 3C~279 implies that most,
or all, of these factors play a major role at different times.

\section*{ACKNOWLEDGEMENTS}
We thank the referee for attentive reading and comments that helped to improve presentation of the manuscript. The data collected by the WEBT collaboration are stored in the WEBT archive at the Osservatorio Astrofisico di Torino - INAF (http://www.oato.inaf.it/blazars/webt/); for questions regarding their availability, please contact the WEBT President Massimo Villata ({\tt massimo.villata@inaf.it}).
The St. Petersburg University team acknowledges support from Russian Scientific Foundation  
grant 17-12-01029. The research at Boston University was supported in part by National Science Foundation grant AST-1615796 and NASA Fermi Guest Investigator grants 80NSSC17K0649, 80NSSC19K1504, and 80NSSC19K1505. The PRISM camera at Lowell Observatory was developed by K. Janes et al. at BU and Lowell Observatory, with funding from the NSF, BU, and Lowell Observatory. The emission-line observations made use of the Discovery Channel Telescope at Lowell Observatory, supported by Discovery Communications, Inc., Boston University, the University of Maryland, the University of Toledo, and Northern Arizona University. The VLBA is an instrument of the National Radio Astronomy Observatory. The National Radio Astronomy Observatory is a facility of the US National
Science Foundation (NSF), operated under cooperative agreement by Associated Universities, Inc. This research has used data from the University of Michigan Radio Astronomy  Observatory which was supported by the  University of Michigan; research at this facility was supported by NASA under awards NNX09AU16G, NNX10AP16G, NNX11AO13G, and NNX13AP18G,  and by the NSF under award AST-0607523. The Steward Observatory spectropolarimetric monitoring project  was supported by NASA Fermi Guest Investigator grants NNX08AW56G, NNX09AU10G, NNX12AO93G, and NNX15AU81G. The Torino group acknowledges financial contribution from agreement ASI-INAF n.2017-14-H.0 and from contract PRIN-SKA-CTA-INAF 2016.
I.A. acknowledges support by a Ram\'on y Cajal grant (RYC-2013-14511) of the "Ministerio de Ciencia, Innovaci\'on, y Universidades (MICIU)" of Spain and from MCIU through the “Center of Excellence Severo Ochoa” award for the Instituto de Astrof\'isica de Andaluc\'ia-CSIC (SEV-2017-0709). Acquisition and reduction of the POLAMI and MAPCAT data were supported by MICIU through grant AYA2016-80889-P. The POLAMI observations were carried out at the IRAM 30m Telescope, supported by INSU/CNRS (France), MPG (Germany) and IGN (Spain). The MAPCAT observations were carried out at the German-Spanish Calar Alto Observatory, jointly operated by the Max-Plank-Institut f\"ur Astronomie and the Instituto de Astrof\'isica de Andaluc\'ia-CSIC. The study is based partly on data obtained with the STELLA robotic telescopes in Tenerife, an AIP facility jointly operated by AIP and IAC.
The OVRO 40-m monitoring program is supported in part by NASA grants NNX08AW31G, NNX11A043G and NNX14AQ89G, and NSF grants AST-0808050 and AST-1109911. T.H. was supported by the Academy of Finland projects 317383 and 320085.
AZT-24 observations were made within an agreement between  Pulkovo, Rome and Teramo observatories. 
The Submillimeter Array is a joint project between the Smithsonian Astrophysical Observatory and the Academia Sinica Institute of Astronomy and Astrophysics and is funded by the Smithsonian Institution and the Academia Sinica.
The Abastumani team acknowledges financial support by the
Shota Rustaveli National Science Foundation under contract
FR/217950/16.  This research was partially supported by the Bulgarian National Science
Fund of the Ministry of Education and Science under grants DN 08-1/2016,
DN 18-13/2017, KP-06-H28/3 (2018) and KP-06-PN38/1 (2019), Bulgarian
National Science Programme ”Young Scientists and Postdoctoral Students
2019”, Bulgarian National Science Fund under grant DN18-10/2017 and
National RI Roadmap Projects DO1-157/28.08.2018 and DO1-153/28.08.2018 of
the Ministry of Education and Science of the Republic of Bulgaria. GD and OV gratefully acknowledge observing grant support from the
Institute of Astronomy and Rozhen National Astronomical Observatory via bilateral joint research project "Study of ICRF radio-sources and fast
variable astronomical objects" (head - G.Damljanovic). This work was partly supported by the National Science Fund of the
Ministry of Education and Science of Bulgaria under grant DN 08-20/2016,
and by project RD-08-37/2019 of the University of Shumen. This work is a part of Projects
No. 176011,
No. 176004,
and No. 176021,
supported by the Ministry of Education, Science and Technological
Development of the Republic of Serbia. M.G.\ Mingaliev acknowledges support through the Russian Government Program of Competitive Growth of Kazan Federal University. The Astronomical Observatory of the Autonomous Region of the Aosta Valley (OAVdA) is managed by the Fondazione Cl\'ement Fillietroz-ONLUS, which is supported by the Regional Government of the Aosta Valley, the Town Municipality of Nus and the "Unit\'e des Communes vald\^otaines Mont-\'Emilius". The research at the OAVdA was partially funded by several "Research and Education" annual grants from Fondazione CRT. 
This article is partly based on observations made with the IAC80
and TCS telescopes operated by the Instituto de Astrof\'isica de Canarias in the Spanish
Observatorio del Teide on the island of Tenerife. 
A part of the observations were carried out using the RATAN-600 scientific equipment (Special Astrophysical Observatory of the Russian Academy of Sciences).

\bibliographystyle{mnras} 
\bibliography{3c279}

\vspace{0.5cm}\noindent
{\large \bf AFFILIATIONS}

\noindent
{\it
$^{1}$Astron. Inst., St.-Petersburg State Univ., Russia\\
$^{2}$Pulkovo Observatory, St.-Petersburg, Russia\\
$^{3}$Institute for Astrophysical Research, Boston University, MA, USA\\
$^{4}$INAF, Osservatorio Astrofisico di Torino, Italy\\
$^{5}$Steward Observatory, University of Arizona, Tucson, AZ85721, USA\\
$^{6}$
Instituto de Astrof\'{\i}sica de Andaluc\'{\i}a-CSIC, Glorieta de la Astronom\'{\i}a s/n, E--18008 Granada, Spain
\\
$^{7}$Instituto de Astrofisica de Canarias (IAC), La Laguna, Tenerife, Spain\\
$^{8}$
Departamento de Astrofisica, Universidad de La Laguna, La Laguna, Tenerife, Spain
\\
$^{9}$
University of Michigan, 323 West Hall, Dept of Astronomy, Ann Arbor MI, 48109-1107, USA
\\
$^{10}$
Institute of applied astronomy, Russian Academy of Sciences, Kutuzov~ emb.~10 191187~St. Petersburg, Russia
\\
$^{11}$
Institute of Astronomy and National Astronomical Observatory, Bulgarian Academy of Sciences, 72 Tsarigradsko shosse Blvd., 1784 Sofia, Bulgaria
\\
$^{12}$
Astronomical Observatory, Department of Physical Sciences, Earth and Environment - University of Siena, Via Roma 56, 53100 Siena, Italy
\\
$^{13}$Crimean Astrophysical Observatory, Russia\\
$^{14}$Dept. of Astronomy, Faculty of Physics, Sofia University, Bulgaria\\
$^{15}$
Osservatorio Astronomico della Regione Autonoma Valle d’Aosta, I-11020 Nus, Italy
\\
$^{16}$EPT Observatories, Tijarafe, La Palma, Spain\\
$^{17}$INAF, TNG Fundacion Galileo Galilei, La Palma, Spain\\
$^{18}$
Max-Planck-Institut f\"ur Radioastronomie, Auf dem H\"ugel 69, 53121 Bonn, Germany
\\
$^{19}$
Graduate Institute of Astronomy, National Central University, Jhongli City, Taoyuan County 32001, Taiwan
\\
$^{20}$Astronomical Observatory, Volgina 7, 11060 Belgrade, Serbia\\
$^{21}$INAF, Osservatorio Astronomico di Roma, Italy\\
$^{22}$
INAF - Osservatorio Astrofisico di Catania, via S. Sofia, 78, I-95123 Catania, Italy
\\
$^{23}$ASI Italian Space Agency, Italy\\
$^{24}$
Center for Astrophysics $|$ Harvard \& Smithsonian, 60 Garden Street, Cambrige, MA 02138 USA
\\
$^{25}$
Finnish Centre for Astronomy with ESO, University of Turku, FI-20014 University of Turku, Finland
\\
$^{26}$
Aalto University Mets\"ahovi Radio Observatory, FI-02540 Kylm\"al\", Finland\\
$^{27}$
Department of Physics and Astronomy, University of Shumen, 9700 Shumen, Bulgaria
\\
$^{28}$
California Institute of Technology, 1216 E California Blvd, Pasadena, CA 91125, USA
\\
$^{29}$Abastumani Observatory, Mt. Kanobili, 0301 Abastumani, Georgia\\
$^{30}$
Astro Space Center of Lebedev Physical Institute, Profsoyuznaya~St.~84/32, 117997~Moscow, Russia
\\
$^{31}$
Institute of Physics and Technology, Dolgoprudny, Institutsky per., 9, Moscow region, 141700, Russia
\\
$^{32}$
Engelhardt Astronomical Observatory, Kazan Federal University, Tatarstan, Russia
\\
$^{33}$
Landessternwarte, Zentrum f\"ur Astronomie der Universit\"at Heidelberg, 69117 Heidelberg, Germany
\\
$^{34}$
Aalto University Dept of Electronics and Nanoengineering, FI-00076 Aalto, Finland
\\
$^{35}$Astronomical Institute, Osaka Kyoiku University, Osaka, 582-8582, Japan
\\
$^{36}$
Special Astrophysical Observatory, Russian Academy of Sciences, Nizhnii Arkhyz, 369167 Russia
\\
$^{37}$Kazan Federal University, 18 Kremlyovskaya St., Kazan, 420008, Russia\\
$^{38}$
Ulugh Beg Astronomical Institute, Uzbekistan Academy of Sciences, Astronomical street 33, Tashkent 100052, Uzbekistan
\\
$^{39}$
INAF, Istituto di Radioastronomia, via Piero Gobetti 93/2, 40129 Bologna, Italy
\\
$^{40}$Osservatorio Astronomico Sirio, Italy\\
$^{41}$Department of Physics, University of Colorado, Denver, USA\\
$^{42}$
Physical Research Laboratory, Navrangpura, Ahmedabad, Gujarat 380009, India\\
$^{43}$Institute of Astronomy, Bulgarian Academy of Sciences, Sofia, Bulgaria
\\
$^{44}$
Instituto de Radio Astronom\'ia Millim\'etrica, Avenida Divina Pastora, 7, Local 20, E--18012 Granada, Spain
\\
 }

\bsp
\label{lastpage}
\end{document}